\documentclass[12pt]{article}

\usepackage{amssymb}
\usepackage{amsmath}
\usepackage[pdftex]{graphicx}   %For arXiv
\usepackage{amsthm}   %This is necessary for "proof"
\usepackage{authblk}   %This is necessary for author descriptions
\usepackage{lscape}   %This is necessary for "landscape"
\usepackage{comment}
\usepackage{setspace}
\makeatletter

\@addtoreset{equation}{section}
\makeatother

\newtheorem{theo}{Theorem}

\allowdisplaybreaks[4]

\newcommand{\indep}{\mathop{\perp\!\!\!\perp}}

\newcommand{\bld}{\boldsymbol}

\usepackage[top=1in,bottom=1in,left=1in,right=1in]{geometry}

\title{Robust Estimating Method for Propensity Score Models and its Application to Some Causal Estimands: A review and proposal}
\author[1]{Shunichiro Orihara}
\affil[1]{Graduate School of Data Science, Yokohama City University, Kanagawa, Japan}
\date{}

\begin{document}
\begin{singlespace}
\maketitle
\end{singlespace}
\section*{Abstract}
In observational study, the propensity score has the central role to estimate causal effects. Since the propensity score is usually unknown, estimating by appropriate procedures is an indispensable step. A point to note that a causal effect estimator might have some bias if a propensity score model was misspecified; valid model construction is important. To overcome the problem, a variety of interesting methods has been proposed. In this paper, we review four methods: using ordinary logistic regression approach; CBPS proposed by Imai and Ratkovic; boosted CART proposed by McCaffrey and colleagues; a semiparametric strategy proposed by Liu and colleagues. Also, we propose the novel robust two step strategy: estimating each candidate model in the first step and integrating them in the second step. We confirm the performance of these methods through simulation examples by estimating the ATE and ATO proposed by Li and colleagues. From the results of the simulation examples, the boosted CART and CBPS with higher-order balancing condition have good properties; both the estimate of the ATE and ATO has the small variance and the absolute value of bias. The boosted CART and CBPS are useful for a variety of estimands and estimating procedures.

\vspace{0.5cm}
\noindent
{\bf Keywords}: Causal inference, Overlap weight, Propensity score, Robust estimation
\section{Introduction}
To estimate causal effects accurately, adjusting covariates (or confounders; hereafter, ``covariates") is one of the important steps in observational study. When all covariates are observed, the covariates can be adjusted, and an unbiased estimator for causal effects can be obtained; the situation of ``no unmeasured confounding" (Hern\'{a}n and Robins, 2020). In this situation, the propensity score has the central role to estimate causal effects (Rosenbaum and Rubin, 1983). However, the propensity score needs to be estimated by appropriate procedures since it is usually unknown. A point to note that an interested estimator might have some bias if a propensity score model was misspecified (Kang and Schafer, 2007); valid model construction is important.

Recently, some causal estimands, the target population of a causal effects estimation, are considered. One of the well-known estimands is the average treatment effects (ATE; Imbens and Rubin, 2015). The ATE assumes causal effects that the overall population is treated ``counterfactual" study treatment and control treatment. In short, the ATE is interested in the overall causal effects. A different estimand called the average treatment effects for the overlap population (ATO) is proposed by Li et al. (2018). The ATO has several attractive features compared with the ATE (Li et al., 2018); for instance, the ATO may become more reasonable causal effects in the sense that subjects who could change their actual treatment to the ``counterfactual" treatment have larger weights than subjects who could not change, and is less sensitive to extreme propensity scores (i.e. low propensity scores in the treatment group, or high propensity scores in the control group). Another estimand is defined by the incremental propensity score (Kennedy, 2019). The estimands are defined how change in the intervention probability (i.e., the propensity score) affects the causal effects. There are some attractive estimands, however, there is a crucial problem; some estimands depend on the ``true" propensity score. For instance, as mentioned Mao et al. (2019), the ATO has some bias when the propensity score model is misspecified. Therefore, the similar problem mentioned in Kang and Schafer (2007) may be occurred.

There may be a critical problem for not only estimators of causal effects but also estimands itself when the propensity score model is misspecified. To overcome the problem, misspecification of the propensity score model, a variety of interesting methods has been proposed. Covariate balancing methods are fully parametric strategy (Imai and Ratkovic, 2014; Fong et al., 2018). The idea is intuitive and easy to understand; a parametric propensity score model is estimated as the covariate balance between treatment groups is balanced sufficiently. This is somewhat similar procedure as calibration methods (Zubizarreta, 2015; Wang and Zubizarreta, 2020) or multiply robust estimations (Han, 2014; Orihara and Hamada, 2019), however, to my understanding, the philosophy is different; the former is estimating weights nonparametrically so that each group's weighted expectation of outcome model becomes the marginal expectation of each potential outcome, and the latter is estimating weights so that the weighted score statistics for propensity score models and outcome models become $0$. Nonparametric strategy such as machine learning methods is also considered widely (McCaffrey et al., 2004; Lee et al., 2010; Westreich et al., 2010; Cannas and Arpino, 2019). As mentioned in Cannas and Arpino (2019), by using a sufficient complex model, checking the balancing property (Rosenbaum and Rubin, 1983) may not be that much of a concern. Although the strategy of covariate balancing methods and machine learning methods are different, objective is similar. Recently, a method that is an intermediate between parametric and nonparametric methods has been proposed (Liu et al., 2018). Liu et al. (2018) mentioned that the proposed method is semiparametric in the sense that the method has both sides. In the parametric side, it is assumed that (high dimensional) covariates information is explained by (low dimensional) linear combinations of the covariates. This spirit is the same as a sufficient dimension reduction (Li, 1991). In the nonparametric side, it is assumed that the explicit form of link functions is not assumed.

In this paper, we compare the explained three methods: the covariate balancing methods (CBPS), the machine learning methods (boosted CART (bCART) algorithm), and the method proposed in Liu et al. (2018). Also, we propose an another parametric strategy in this paper: a multiply robust estimator for the propensity score (note that ``multiply robust" is slightly different from Han, 2014). In brief, we prepare some candidate models, and combine the candidate models all at once (do not have to select just one); we call the method as ``integrated method". To combine the candidate models, we consider two procedures. One is a well-known model averaging approach (MA; c.f. Hoeting et al., 1999; Hjort and Claeskens, 2003; Xie et al., 2019), the other one is novel which has some good theoretical properties. Note that the former is called ``MA-based integrated method", the latter is ``proposed integrated method". For the proposed integrated method, we confirm theoretical properties. If the correct model was included in the candidate models, the parameter estimator would have the consistency. Also, we consider the situation where the true model is not included in the candidate models. Under this situation, we confirm the condition where the estimator becomes a ``valid" in the sense of the true Kullback-Leibler (KL) divergence; this property is different from MA. By using the integrated methods, we consider a robust estimating method to estimate a ``valid" causal effects without any model/variable selection methods. Therefore, these methods may overcome the problem of a variable selection of a propensity score model (Brookhart et al., 2006; Austin et al., 2007). For instance, a researcher claims that some covariates have to be included in a propensity score model, whereas, another researcher claims that the covariates do not have to be included. Commonly, in this situation, one of the researchers has to compromise, or the researchers use some model/variable selection methods. Meanwhile, integrated procedure provides a simple solution: the both models can be included in a one integrated model. Note that no methods explained here are suffer from the ``propensity score tautology'' (Imai and Ratkovic, 2014), that is, we need not take care of model adjustment to satisfy the ``balancing property'' (Rosenbaum and Rubin, 1983). In this sense, the methods have reasonable to apply real-world data since there may be too many covariates to construct an appropriate propensity score model.

The remainder of the paper proceeds as follows. In section 2, we review the previous methods, and explain the proposed methods briefly. Regarding the proposed integrated method, we confirm the theoretical properties as a new robust generalized linear model (GLM) estimator with and without the condition where the true model is included in the candidate models. Since the propensity score model is assumed as not only a logistic regression model but also a regression model with normal error under general treatment regimes (Hirano and Imbens, 2004 and Imai and van Dyk, 2004), considering the general situation is valuable. In section 3, we show simulation data examples when the ATE and ATO is central interest, respectively. The simulation datasets refers to well-established simlation examples (Setoguchi et al., 2008; Setodji et al.,2017; Huang et al., 2022). Some materials are found in Appendix.

\section{Review of previous methods and propose the novel methods}
Let $n$ be the sample size. $T_{i}\in\{0,\, 1\}$, $\boldsymbol{X}_{i}\in \mathbb{R}^{p}$ and $(Y_{1i},\, Y_{0i})\in \mathcal{Y}\subset\mathbb{R}^{2}$ denote the treatment, a vector of covariates measured prior to treatment, and potential outcomes, respectively. $Y_{i}:=T_{i}Y_{1i}+(1-T_{i})Y_{0i}$ denotes an observed outcome and we assume that $i=1,\, 2,\, \dots,\, n$ are i.i.d. samples and the stable unit treatment value assumption (c.f. Rosenbaum and Rubin, 1983) holds.

Next, we introduce a general class of estimands called ``weighted average treatment effect" (WATE) in Hirano et al. (2003) and Li et al. (2018). The ATE conditional on $\bld{x}$ is defined as $\tau(\bld{x}):={\rm E}[Y_{1}-Y_{0}|\bld{x}]$, and the WATE is defined as
$$
\tau_{h}:=\frac{{\rm E}[\tau(\bld{X})h(\bld{X})]}{{\rm E}[h(\bld{X})]},
$$
where $h(\cdot)$ is a weight function for $\bld{x}$. When $h(\bld{x})\equiv 1$, the WATE becomes the ATE: $\tau_{ATE}:={\rm E}[Y_{1}-Y_{0}]$ obviously. Under the  strongly ignorable treatment assignment assumption $T\indep(Y_{1},\, Y_{0})|\boldsymbol{X}$, we can estimate the ATE using the propensity score $e(\boldsymbol{X}_{i}):={\rm P}(T_{i}=1|\boldsymbol{X}_{i})$. When $h(\bld{x})=e(\bld{x})(1-e(\bld{x}))$, the WATE becomes the ATO (Li et al., 2018):
$$
\tau_{ATO}:=\frac{{\rm E}[\tau(\bld{X})e(\bld{X})(1-e(\bld{X}))]}{{\rm E}[e(\bld{X})(1-e(\bld{X}))]}.
$$
Since the function $e(1-e),\, e\in(0,1)$ is the convex and symmetric function at $1/2$, subjects who could change their actual treatment to the ``counterfactual" treatment (i.e. $e\approx 1/2$) have larger weights than subjects who could not change (i.e. $e\approx 0$ or $1$). 

As mentioned in Introduction, the propensity score is commonly unknown and need to be estimated. If a propensity score model is misspecified, an estimated ATE may have serious bias. Also, as obviously, the ATO depends on the true propensity score; if the estimated propensity score model is misspecified, the ATO has some bias (see also Mao et al., 2019). In the following subsections, we introduce and propose robust propensity score estimating methods.
\subsection{Covariate balancing propensity score}
Assuming that a propensity score model has a parametric model $e(\bld{X};\bld{\beta})$ with $q$-dimensional parameter $\bld{\beta}$. Commonly, the model is assmed as the logistic regression model:
$$
e(\bld{X};\bld{\beta})=\frac{\exp\left\{\bld{X}^{\top}\bld{\beta}\right\}}{1+\exp\left\{\bld{X}^{\top}\bld{\beta}\right\}}.
$$
Under the model, the parameter $\bld{\beta}$ is estimated by the following estimating equation:
\begin{align}
\label{CBPS_1}
\sum_{i=1}^{n}\frac{T_{i}g(\bld{X}_{i})}{e(\bld{X}_{i};\bld{\beta})}=\sum_{i=1}^{n}\frac{(1-T_{i})g(\bld{X}_{i})}{1-e(\bld{X}_{i};\bld{\beta})},
\end{align}
where $g(\bld{X}_{i})$ is some measurable $r$-dimensional function. Note that $r\geq q$; this is the situation of the ``overidentified restrictions''. Therefore, $\bld{\beta}$ is solved as the procedure of the generalized method of moments (see Imai and Ratkovic, 2014). For instance, $g(\bld{X}_{i})$ is assumed as first- and second-order moments of the covariate $\bld{X}$: 
\begin{align}
\label{CBPS_2}
g(\bld{X}_{i})=\left(
X_{i1},\dots,X_{ip},X^2_{i1},\dots,X^2_{ip}
\right).
\end{align}
However, it is under discussion that how many order moments of the covariate need to be included in. Huang et al. (2022) mentioned that by including the third-order moments, the absolute bias of the interested causal effects can be reduced greatly when there is nonlinear relationships between a treatment and covariates, and an outcome and covariates.

By the solution of (\ref{CBPS_1}) $\hat{\bld{\beta}}$ is substituted in the model, the covariate balancing propensity score (CBPS) is obtained: $e(\bld{X};\hat{\bld{\beta}})$. The CBPS has the reasonable property: the balancing property. This is straightforward from (\ref{CBPS_1}) under the true propensity score:
\begin{align*}
{\rm E}\left[\frac{Tg^{*}(\bld{X})}{e(\bld{X})}\right]&={\rm E}_{\bld{X}}\left[\frac{g^{*}(\bld{X})}{e(\bld{X})}{\rm E}\left[T|\bld{X}\right]\right]={\rm E}_{\bld{X}}\left[g^{*}(\bld{X})\right]\\
{\rm E}\left[\frac{(1-T)g^{*}(\bld{X})}{1-e(\bld{X})}\right]&={\rm E}_{\bld{X}}\left[\frac{g^{*}(\bld{X})}{1-e(\bld{X})}\left(1-{\rm E}\left[T|\bld{X}\right]\right)\right]={\rm E}_{\bld{X}}\left[g^{*}(\bld{X})\right].
\end{align*}
Note that the above equation is hold under all measurable functions $g^{*}(\bld{X}_{i})$. Therefore, regarding the CBPS, the balancing property is hold partially (more precisely, first- and second-order moments of the covariate $\bld{X}$ under (\ref{CBPS_2})). In this sense, it is appeared that the selection of (\ref{CBPS_2}) is critical when using the CBPS.

The CBPS is applied easily by ``CBPS'' function of CBPS library in R (Fong et al., 2022). Note that the default of the CBPS function assumes the estimand as ATT. By using the option ``ATT=0'', the CBPS satisfying the condition (\ref{CBPS_1}) is obtained.

\subsection{boosted CART}
The bCART algorithm is initially proposed by McCaffrey et al. (2004), and their algorithm is justified as woking well by the following researches (e.g. Lee et al., 2010; Westreich et al., 2010). Assuming that the propensity score has the following formulation:
$$
e(\bld{X})=\frac{\exp\left\{\eta\left(\bld{X}\right)\right\}}{1+\exp\left\{\eta\left(\bld{X}\right)\right\}},
$$
where $\eta(\cdot)$ is a smoothed function. To estimate the propensity score, the (sample) expectation of the log likelihood function is considered:
\begin{align}
\label{bct_1}
{\rm E}\left[\ell(\eta)\right]={\rm E}\left[T\eta\left(\bld{X}\right)-\log\left\{1+\exp\left\{\eta\left(\bld{X}\right)\right\}\right\}|\bld{X}\right],
\end{align}
and considering maximization of (\ref{bct_1}) regarding $\eta(\cdot)$; in this sense, bCART is a likelihood-based method. To update the function (\ref{bct_1}), we would like to find the function $h(\bld{X})$ such that
$$
{\rm E}\left[\ell(\hat{\eta}+\lambda h)\right]>{\rm E}\left[\ell(\hat{\eta})\right]
$$
for some $\lambda$. Namely, searching the direction of the top of the likelihood function. By the result of Friedman (2001), the ``best'' function of $h(\bld{X})$ can be derived as
$$
h(\bld{X})={\rm E}\left[T-e(\bld{X})|\bld{X}\right].
$$
To estimate $h(\bld{X})$, some nonparametric procedures such as regression tree algorithm are applied. Note that the initial value of $\hat{\eta}$ is recommended as $\hat{\eta}=\log\{\bar{T}/(1-\bar{T})\}$. More detail of the algorithm is explained from the page 34 to the page 36 in McCaffrey et al. (2004).

The bCART is applied easily by ``ps'' function of twang library in R (Ridgeway et al., 2015).

\subsection{Liu et al. (2018)}
Assuming that the true propensity score has the following formulation:
\begin{align}
\label{Liu_1}
e(\bld{X};\bld{\beta})=\frac{\exp\left\{\eta\left(\bld{X}^{\top}\bld{\beta}\right)\right\}}{1+\exp\left\{\eta\left(\bld{X}^{\top}\bld{\beta}\right)\right\}},
\end{align}
where $\eta(\cdot)$ is an arbitaly function, and $\bld{\beta}$ is a $p\times q$ matrix. Note that the above ``logistic form'' is not essential but to transform into the range of $(0,1)$ since the arbitaly form of $\eta(\cdot)$; this is the nonparametric side of Liu et al. (2018) mentioned in Introduction. Additionally, we assume that the propensity score is explained by a $q$-dimensional linear combination of the covariates $\bld{X}^{\top}\bld{\beta}$. In other words, the propensity score is explained by the $q$-dimensional parameters; this is the same spirit as a sufficient dimension reduction (Li, 1991), and  this is the parametric side of the semiparametric strategy.

Unfortunately, we cannot estimate (\ref{Liu_1}) directly since $\eta(\cdot)$ is an arbitary function. To solve the problem, Liu et al. (2018) propose the two step estimating procedure. In the first step, the initial estimator $\tilde{\bld{\beta}}$ is obtained from an estimating equation under $\eta\left(\bld{X}^{\top}\bld{\beta}\right)=\bld{1}_{d}^{\top}\bld{X}^{\top}\bld{\beta}$:
\begin{align}
\label{Liu_2}
\sum_{i=1}^{n}{\rm vecl}\left[\left(\bld{X}_{i}-\widehat{\rm E}\left[\bld{X}|\bld{X}_{i}^{\top}\bld{\beta}\right]\right)\left(T_{i}-\frac{\exp\left\{\bld{1}_{q}^{\top}\bld{X}_{i}^{\top}\bld{\beta}\right\}}{1+\exp\left\{\bld{1}_{q}^{\top}\bld{X}_{i}^{\top}\bld{\beta}\right\}}\right)\bld{1}_{q}^{\top}\right]=\bld{0},
\end{align}
where ${\rm vecl}[\cdot]$ is some ``lower part'' of elements after the vectorization (see Liu et al., 2018), and $\widehat{\rm E}\left[\bld{X}|\bld{X}_{i}^{\top}\bld{\beta}\right]$ is a Nadaraya-Watson kernel estimator. Note that the form $\eta\left(\bld{X}^{\top}\bld{\beta}\right)=\bld{1}_{d}^{\top}\bld{X}^{\top}\bld{\beta}$ is not essential; More arbitary forms are considered in Ghosh et al. (2021). From (\ref{Liu_2}), the local efficient estimator $\tilde{\bld{\beta}}$ is obtained; it is used as the initial value of the following step. In the second step, to estimate the primary estimator of $\bld{\beta}$, a kernel approximation of $\eta$ and $\bld{\eta}'$ is implemented, where $\bld{\eta}'$ is the first derivative of $\eta$:
$$
\sum_{i=1}^n\left[T_i-\frac{\exp \left\{b_0+\bld{b}_1^{\top}\left(\bld{X}_{i}^{\top}\bld{\beta}-\bld{x}_{0}^{\top}\bld{\beta}\right)\right\}}{1+\exp \left\{b_0+\bld{b}_1^{\top}\left(\bld{X}_{i}^{\top}\bld{\beta}-\bld{x}_{0}^{\top}\bld{\beta}\right)\right\}}\right] K_h\left(\bld{X}_{i}^{\top}\bld{\beta}-\bld{x}_{0}^{\top}\bld{\beta}\right)=0,
$$
$$
\sum_{i=1}^n\left[T_i-\frac{\exp \left\{b_0+\bld{b}_1^{\top}\left(\bld{X}_{i}^{\top}\bld{\beta}-\bld{x}_{0}^{\top}\bld{\beta}\right)\right\}}{1+\exp \left\{b_0+\bld{b}_1^{\top}\left(\bld{X}_{i}^{\top}\bld{\beta}-\bld{x}_{0}^{\top}\bld{\beta}\right)\right\}}\right]\left(\bld{X}_{i}^{\top}\bld{\beta}-\bld{x}_{0}^{\top}\bld{\beta}\right) K_h\left(\bld{X}_{i}^{\top}\bld{\beta}-\bld{x}_{0}^{\top}\bld{\beta}\right)=\bld{0},
$$
where $K_h$ is some kernel function. $\hat{b}_0$ and $\hat{\bld{b}}_1$, the solution of the above formulae at $\bld{x}_{0}^{\top}\bld{\beta}$, are the estimator of $\eta$ and $\bld{\eta}'$, respectively. By using the estimators $\tilde{\bld{\beta}}$, $\hat{b}_0$, and $\hat{\bld{b}}_1$, the solution of the following efficient score becomes the target efficient estimator $\hat{\bld{\beta}}$:
\begin{align*}
\sum_{i=1}^{n}\bld{S}_{\rm eff}\left(T_i,\, \bld{X}_i,\, \bld{X}_{i}^{\top}\bld{\beta},\, \hat{\rm E},\, \hat{\eta},\, \hat{\bld{\eta}}'\right)&\\
&\hspace{-4.5cm}=\sum_{i=1}^{n}{\rm vecl}\left[\left\{\bld{X}_i-\hat{\rm E}\left(\bld{X}_i|\bld{X}_{i}^{\top}\bld{\beta}\right)\right\}\left(T_i-\frac{\exp \left\{\hat{\eta}\left(\bld{X}_{i}^{\top}\bld{\beta}\right)\right\}}{1+\exp \left\{\hat{\eta}\left(\bld{X}_{i}^{\top}\bld{\beta}\right)\right\}}\right) \hat{\bld{\eta}}'\left(\bld{X}_{i}^{\top}\bld{\beta}\right)^{\top}\right].
\end{align*}
Note that the above steps are justified from some viewpoints (Ma and Zhu, 2012). Also, this estimating procedure are expanded to an augmented IPW estimator (AIPW; Ghosh et al., 2021).

The R package for the above procedure including the AIPW estimator is released as ``SDRcausal" package developed by stat4reg team (Ghasempour and de Luna, 2021). Procedure of package installing is described in the page 2 of Ghasempour and de Luna (2021). By using ``ipw.ate" function, the above procedure can be implemented.

\subsection{Proposed methods}
We propose novel parametric strategies which integrate some candidate parametric propensity score models. The proposed strategies have two steps. At the first step, assuming the $K$ parametric models:
$$
f(t|\bld{x};\bld{\beta}_{k})=\exp\left\{t\theta_{k}(\bld{x};\bld{\beta}_{k})+\log\left(1+e^{\theta_{k}(\bld{x};\bld{\beta}_{k})}\right)\right\},
$$
where $\theta=\log\left(\frac{p}{1-p}\right)$. When we assume the logistic regression model, $p(\bld{x};\bld{\beta})=\frac{\exp\left\{\bld{x}^{\top}\bld{\beta}\right\}}{1+\exp\left\{\bld{x}^{\top}\bld{\beta}\right\}}$. Also, when we assume the probit model, $p(\bld{x};\bld{\beta})=\Phi\left(\bld{x}^{\top}\bld{\beta}\right)$. For each model $k$, ordinary estimating procedures such as MLE is applied, and each estimator $\hat{\bld{\beta}}_{k}$ is derived.

At the second step, two options can be considered. First, an ordinary model averaging strategy can be applied. Concretely, the estimated propensity score by MA-based integrated procedure becomes
$$
e\left(\bld{x};\hat{\bld{\beta}}_{1},\dots,\hat{\bld{\beta}}_{K}\right)=\sum_{k=1}^{K}w_{k}f(t=1|\bld{x};\hat{\bld{\beta}}_{k}).
$$
$w_{k}$ are some weights such as derived from an information criterion:
$$
w_{k}=\frac{\exp\left\{-\frac{IC_{k}}{2}\right\}}{\sum_{k'=1}^{K}\exp\left\{-\frac{IC_{k'}}{2}\right\}},
$$
where $IC_{k}$ is a value of an information criterion of model $k'$ such as AIC and BIC.

Secondly, we propose the novel strategy which has different properties from MA. Concretely, the integrated likelihood of $K$ models becomes
\begin{align}
\label{NV_1}
f\left(t|\bld{x};\bld{\gamma},\hat{\bld{\beta}}_{1},\dots,\hat{\bld{\beta}}_{K}\right)=\exp\left\{t\sum_{k=1}^{K}\gamma_{k}\theta_{k}(\bld{x};\hat{\bld{\beta}}_{k})-\log\left(1+e^{\sum_{k=1}^{K}\gamma_{k}\theta_{k}(\bld{x};\hat{\bld{\beta}}_{k})}\right)\right\}.
\end{align}
By estimating $\bld{\gamma}$ such as MLE, the estimated propensity score by proposed integrated method becomes
$$
e\left(\bld{x};\hat{\bld{\gamma}},\hat{\bld{\beta}}_{1},\dots,\hat{\bld{\beta}}_{K}\right)=f\left(t=1|\bld{x};\hat{\bld{\gamma}},\hat{\bld{\beta}}_{1},\dots,\hat{\bld{\beta}}_{K}\right).
$$
Theoretical properties under the GLM setting are appeared in Appendix \ref{appA}.

One of the strong points of the proposed methods is ease to use; only related to GLM procedure is used. It relates to ``glm" function in R, and ``GENMOD" procedure in SAS. Specifically, it may be important for drug development situation to use the method by SAS simply.

\section{Simulation data examples}
As mentioned in the previous section, there are many robust methods for propensity score model misspecification. In this section, we confirm their performance through simulation data examples. About the example data set, we confirm the performance when we are interested in the ATE and ATO, respectively. Note that an iteration time of all simulations is 500. Some simulation results are appeared in Appendix \ref{appC}.

At first, we explain the data generating mechanism in this simulation. The setting is referering to Wyss et al. (2014); Setodji et al. (2017); Mao et al. (2019); Huang et al. (2022). There are 10 covariates generated from the following distributions:
\begin{align*}
X_{i1},\ X_{i3}&\stackrel{i.i.d.}{\sim}Ber(0.5),\\
\left(X_{i2},X_{i4},X_{i7},X_{i10}\right)^{\top}&\stackrel{i.i.d.}{\sim}N(\bld{0}_{4},{\rm I}_{4}),
\end{align*}
$$
X_{i5}\sim Ber(p_{i5}),\ X_{i6}\sim Ber(p_{i6}),\ X_{i8}\sim Ber(p_{i8}),\ X_{i9}\sim Ber(p_{i9}),
$$
where
$$
p_{i5}=expit\left\{0.4(X_{i1}-1)\right\},\ p_{i6}=expit\left\{5X_{i2}\right\},\ p_{i8}=expit\left\{0.4(X_{i3}-1)\right\},\ p_{i9}=expit\left\{5X_{i4}\right\}.
$$
From the setting, the correlations between the covariates become approximately as follows:
$$
Corr(X_{1},X_{5})=Corr(X_{3},X_{8})=0.2,\ Corr(X_{2},X_{6})=Corr(X_{4},X_{9})=0.8.
$$
Note that the correlations of the other combinations are approximately $0$. Next, we introduce the true propensity score model. The model is the same as Option E of Setodji et al. (2017):
$$
e(\bld{X}_{i})={\rm P}\left(T=1|\bld{X}_{i}\right)=expit\left\{h_{T}^{\top}\left(\bld{X}_{i}\right)\bld{\beta}\right\},
$$
where $\beta_{k}\sim U(-0.8,0.8)$ ($k=1,2,\dots,12$), and
\begin{align*}
h_{T}^{\top}\left(\bld{X}_{i}\right)&=\left(X_{i1},\dots,X_{i7},X_{i1}\times X_{i3},X_{i2}\times X_{i4},X_{i4}\times X_{i5},X_{i5}\times X_{i6},X_{i2}^2\right).
\end{align*}
Therefore, there is obviously nonlinear structure in the propensity score; the standard linear logistic regression may have some bias. Finally, we introduce the true outcome model:
$$
Y_{i}=\left(0.5+X_{i2}+X_{i1}\times X_{i2}\right)T_{i}+h_{Y}^{\top}\left(\bld{X}_{i}\right)\bld{\xi}+\varepsilon_{i},\ \ \ \varepsilon_{i}\stackrel{i.i.d.}{\sim}N(0,1),
$$
where $\xi_{k}\sim U(-0.73,0.73)$ ($k=1,2,\dots,8$), and
$$
h_{Y}^{\top}\left(\bld{X}_{i}\right)=\left(1,\exp\left\{X_{i1}\right\},\exp\left\{X_{i2}\right\},\exp\left\{X_{i3}\right\},\exp\left\{1.3X_{i4}\right\},X_{i8},X_{i9},X_{i10}\right).
$$
Therefore, there are the heterogeneous treatment effects, and nonlinear structure in the outcome model.

From the settings, $X_{i1}$, $X_{i2}$, $X_{i3}$, and $X_{i4}$ are confounders; it is at least necessary to adjust the covariates for estimating unbiased causal effects. The true causal effect for the ATE is approximately 0.500, and for the ATO is approximately 0.430; the methods which estimate the true causal effect acculately are more appropriate.

\subsection{Application to the ATE}
To estimate the ATE, we consider the IPW-type estimator:
$$
\hat{\tau}^{ipw}_{ATE}:=\frac{\sum_{i=1}^{n}\frac{T_{i}Y_{i}}{\hat{e}_{i}}}{\sum_{i=1}^{n}\frac{T_{i}}{\hat{e}_{i}}}-\frac{\sum_{i=1}^{n}\frac{(1-T_{i})Y_{i}}{1-\hat{e}_{i}}}{\sum_{i=1}^{n}\frac{1-T_{i}}{1-\hat{e}_{i}}},
$$
where $\hat{e}_{i}$ is any propensity score estimate. To estimate the propensity score, we consider the following estimating methods.
\subsubsection{Logistic regression}
Considering a linear logistic regression model:
$$
{\rm P}\left(T=1|\bld{X}_{i};\bld{\beta}\right)=expit\left\{\left(X_{i1},X_{i2},\dots,X_{i10}\right)\bld{\beta}\right\}.
$$
Since there is only linear structure, the estimated propensity score has some bias obviously. In the following simulation studies, we use the ``glm'' function with no additional option.
\subsubsection{CBPS}
Regarding the estimating equation (\ref{CBPS_1}), we consider
\begin{align}
\label{sim1}
g(\bld{X}_{i})=\left(
X_{i1},X_{i2},\dots,X_{i4},X^2_{i2},X^2_{i4}
\right).
\end{align}
Therefore, we consider the first- and second order balancing conditions for the confounders. In the following simulation studies, we use the ``CBPS'' function with the following options:
\begin{itemize}
\item {\it ATT = 0}; this option is already explained.
\item {\it method = ``exact''}; the likelihood is not used; only balancing condition is used when estimating a parameter of the propensity score model.
\end{itemize}
\subsubsection{bCART}
As CBPS, only confounders: $(X_{i1},X_{i2},\dots,X_{i4})$ are included in the propensity score model. In the following simulation studies, we use the ``ps'' function with the following option:
\begin{itemize}
\item {\it stop.method = ``ks.mean''}; the Kolmogorov-Smirnov statistic is used for decision whether the estimated parameters converge sufficiently.
\end{itemize}
\subsubsection{Liu et al. (2018)}
As CBPS and bCART, the confounders and an intercept term are included in the propensity score model (i.e., $p=5$). Also, we consider that the dimension reduce to 2 dimension: $q=2$.  In the following simulation studies, we use the ``ipw.ate'' function with the following option:
\begin{itemize}
\item {\it alpha\_initial = c(c(1,-0.5+0.5*runif(1*4)),c(1,-0.5+0.5*runif(1*4)))}; the initial value is selected randomly; $\eta\left(\bld{X}^{\top}\bld{\beta}\right)=\bld{Z}^{\top}\left(\bld{X}^{\top}\bld{\beta}\right)$, where $Z_{jk}\sim U(0,1)$, $j=1,2,\dots,5$, $k=1,2$.
\end{itemize}
\subsubsection{Integrated methods}
We consider 3 propensity score models:
\begin{align*}
{\rm P}\left(T=1|\bld{X}_{i};\bld{\beta}_{1}\right)&=expit\left\{\left(X_{i2},X_{i3},X_{i4}\right)\bld{\beta}_{1}\right\},\\
{\rm P}\left(T=1|\bld{X}_{i};\bld{\beta}_{2}\right)&=expit\left\{\left(X_{i1},X_{i2},X_{i1}\times X_{i2}\right)\bld{\beta}_{2}\right\},\\
{\rm P}\left(T=1|\bld{X}_{i};\bld{\beta}_{3}\right)&=expit\left\{\left(X_{i2},X_{i4},X_{i2}\times X_{i4}\right)\bld{\beta}_{3}\right\}.
\end{align*}
For the MA-based methos, the BIC is used to estimate the weights $w_{k}$. As logistic regression, we use the ``glm'' function with no additional option in the following simulation studies.

\begin{landscape}
\begin{figure}[h]
\begin{center}
\begin{tabular}{c}
\includegraphics[width=24cm]{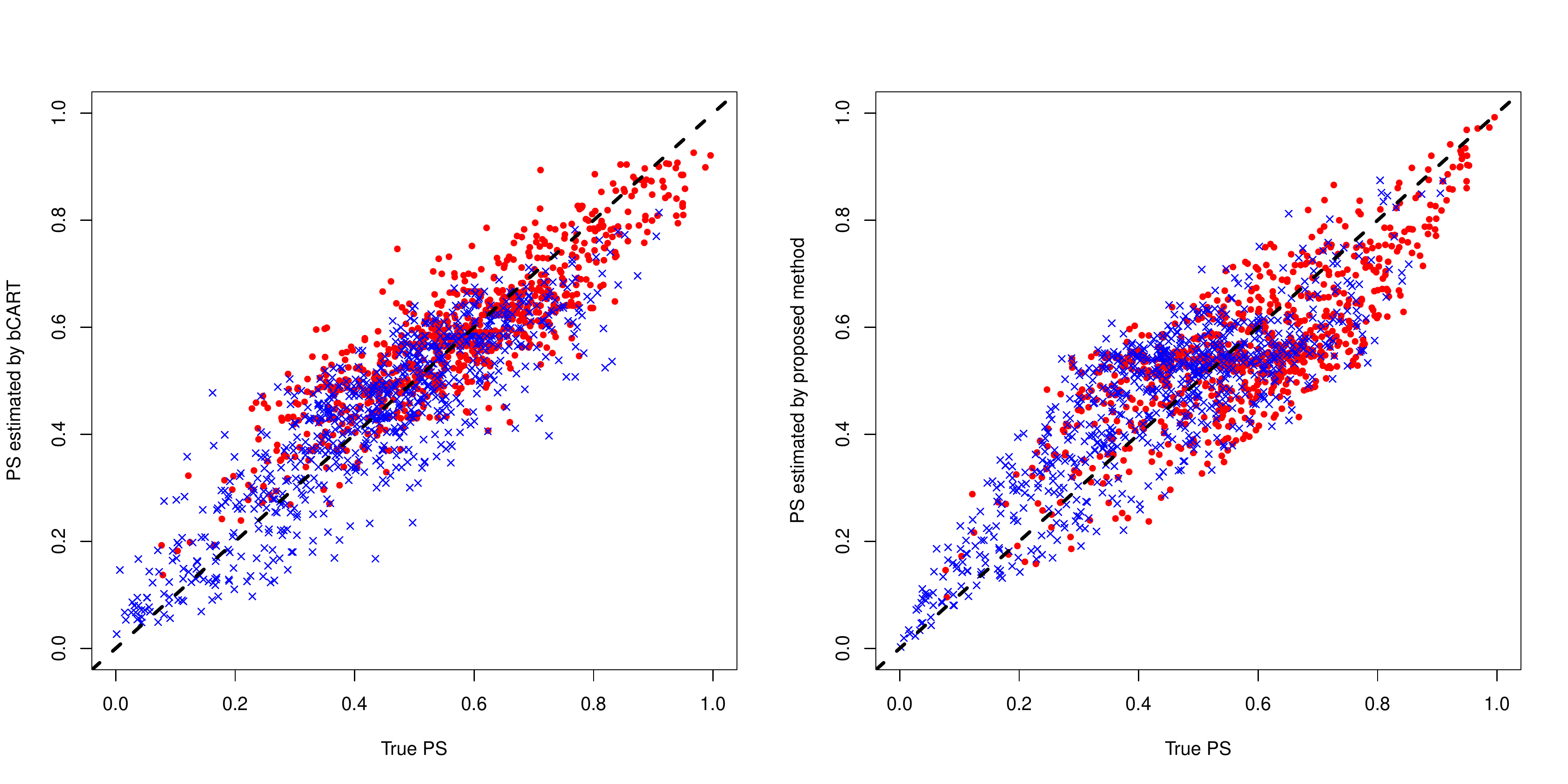}
\end{tabular}
\caption{Scatter plots of each estimating methods in an iteration}
\label{fig8}
\end{center}
{\footnotesize
\begin{description}
\item{\bf Note 1:} Red circle is the treatment group ($T=1$) and the blue cross is the control group ($T=0$).
\item{\bf Note 2:} The scatter plots of the other methods are in Appendix \ref{appC}.
\end{description}
}
\end{figure}
\end{landscape}

The smiulation results are summarized in the Table \ref{tab1}, Figure \ref{fig1} and \ref{fig3}.
\begin{itemize}
\item Logistic regression\\
As expected, there are some remained bias, and the absolute value of bias is the largest in all methods. However, the RMSE is somewhat small compared with the other methods; this is derived from the small variance of the IPW estimator. In this sense, the IPW estimator is stable in this simulation setting.
\item CBPS\\
The direction of bias is the same as the logistic regression result; however the absolute value of the bias is smaller than the logistic regression. This is derived from the covariate balancing condition (\ref{sim1}).  In the large sample situation, the variance of the IPW estimator becomes small, however, the bias is still remained. To confirm the cause of the problem, we consider the additional simulation in the large sample sitruation; using the following moment condition:
$$
g(\bld{X}_{i})=\left(
X_{i1},X_{i2},\dots,X_{i4},X^2_{i2},X^2_{i4},X^3_{i2},X^3_{i4},X_{i1}\times X_{i2}
\right).
$$
The summary of the IPW estimator is in the last row of the Table \ref{tab1}; The IPW estimator has the smallest variance and bias in all methods. From the result, higher-order moment conditions such as the third-order should be included in the balancing condition, as mentioned in Huang et al. (2022).

\item bCART\\
The IPW estimator has the smallest variance in all methods, and relatively small bias in this simulation setting. Surprisingly, the property is hold in the small sample situation despite of the nonparametric strategy. The scatter plot of the true propensity score and estimated one in an iteration is in Figure \ref{fig8} (the left panel). Both the treatment and control group, the propensity score is well estimated, and ``extreme'' propensity scores are very few. In other words, values under the dash line are very few when the true propensity score value is small in the treatment group, and values over the dash line are very few when the true propensity score value is large in the control group. This autometed ``stable'' estimation may derive the stable estimation of the IPW estimator.
\item Liu et al. (2018)\\
The IPW estimator has somewhat small variance and bias in the small sample situation, however, the variance and bias is not resolved in the large sample situation. This is because more than half of iterations, the Liu's method cannot estimate the propensity score properly (see the Note 1 of Table \ref{tab1}). The Liu's method has somewhat good properties, however, it may become unstable. From the result, we do not consider the method in the subsequent simulation.
\item Proposed integrated method\\
The IPW estimator has the smallest bias in all methods, however, the variance is the largest. The scatter plot of the true propensity score and estimated one in an iteration is in Figure \ref{fig8} (the right panel). As bCART, the propensity score is well estimated. However, there are few extreme propensity scores; the propensity scores may cause unstableness of the IPW estimator. Of course, the result is strongly related to the candidate models; we need to consider each propensity score model carefully.
\item MA-based integrated method\\
The properties are very similar as the proposed method. However, the IPW estimator cannot be calculated in the large sample situation (see the Note 2 of Table \ref{tab1}). From the result, we do not consider the method in the subsequent simulation.
\end{itemize}

%\begin{landscape}
\begin{table}[h]
\begin{center}
\caption{Summary of estimates for the ATE}
\label{tab1}
\begin{tabular}{|c||c|c|c|c|}\hline
{\bf Method}&\multicolumn{4}{|c|}{Small sample $n=800$}\\\cline{2-5}
&{Mean (SD)}&{Median (Range)}&Bias&RMSE\\\hline
{\bf Logis. reg.}&0.366 (0.247)&0.383 (-1.55, 1.07)&-0.134&0.281\\\hline
{\bf CBPS}&0.382 (0.180)&0.385 (-0.86, 0.88)&-0.118&0.215\\\hline
{\bf bCART}&0.617 (0.176)&0.618 (-0.22, 1.25)&0.117&0.211\\\hline
{\bf Liu et al.}&0.389 (0.221)&0.427 (-0.48, 0.93)&-0.111&0.247\\\hline
{\bf Proposed}&0.524 (0.513)&0.558 (-6.51, 4.49)&0.024&0.514\\\hline
{\bf MA-based}&0.523 (0.513)&0.558 (-6.51, 4.49)&0.023&0.514\\\hline\hline
{\bf Method}&\multicolumn{4}{|c|}{Large sample $n=1600$}\\\cline{2-5}
&{Mean (SD)}&{Median (Range)}&Bias&RMSE\\\hline
{\bf Logis. reg.}&0.377 (0.178)&0.395 (-0.57, 0.90)&-0.123&0.216\\\hline
{\bf CBPS}&0.383 (0.128)&0.394 (-0.18, 0.73)&-0.117&0.174\\\hline
{\bf bCART}&0.596 (0.117)&0.594 (0.24, 1.00)&0.096&0.152\\\hline
{\bf Liu et al.}&0.328 (0.202)&0.357 (-0.35, 0.68)&-0.172&0.266\\\hline
{\bf Proposed}&0.517 (0.344)&0.552 (-3.37, 1.28)&0.017&0.344\\\hline
{\bf MA-based}&--&--&--&--\\\hline
\begin{tabular}{c}{\bf CBPS}\\{\bf (additional)}\end{tabular}&0.441 (0.092)&0.442 (-0.04, 0.85)&-0.059&0.109\\\hline
\end{tabular}
\end{center}
{\footnotesize
\begin{description}
\item{\bf Note 1:} Some propoensity score estimaties of Liu's method become ``NA''. The number that at reast one propensity score estimate becomes ``NA'' is 327 (65.4\%) in $n=800$, and 386 (77.2\%) in $n=1600$, respectively. The statistics summarized in Table \ref{tab1} are calculated from the iterations there is no ``NA'' values (i.e., $500-327=173$ in $n=800$, and $500-386=114$ in $n=1600$).
\item{\bf Note 2:} The weights $w$ of MA-based method become $0$ in $n=1600$. This is because $\exp\left\{-\frac{IC_{k}}{2}\right\}$ becomes very small value, and is handled as $0$ in R. Therefore, the estimates of MA-based cannot be calculated.
\end{description}
}
\end{table}
%\end{landscape}

\subsection{Application to the ATO}
To estimate the ATO, we consider three estimators: the IPW-type, the AIPW-type, and Bang \& Robins (BR)-type estimator. 
\begin{itemize}
\item IPW-type estimator (Li et al., 2018)
$$
\hat{\tau}^{ipw}_{ATO}:=\frac{\sum_{i=1}^{n}T_{i}(1-\hat{e}_{i})Y_{i}}{\sum_{i=1}^{n}T_{i}(1-\hat{e}_{i})}-\frac{\sum_{i=1}^{n}(1-T_{i})\hat{e}_{i}Y_{i}}{\sum_{i=1}^{n}(1-T_{i})\hat{e}_{i}}.
$$
\item AIPW-type estimator (Mao et al., 2019)
\begin{align*}
\hat{\tau}^{aug}_{ATO}&:=\frac{\sum_{i=1}^{n}\hat{e}_{i}(1-\hat{e}_{i})(\check{m}(1,\bld{X}_{i};\hat{\bld{\xi}})-\check{m}(0,\bld{X}_{i};\hat{\bld{\xi}}))}{\sum_{i=1}^{n}\hat{e}_{i}(1-\hat{e}_{i})}\nonumber\\
&\hspace{0.5cm}+\frac{\sum_{i=1}^{n}T_{i}(1-\hat{e}_{i})(Y_{i}-\check{m}(1,\bld{X}_{i};\hat{\bld{\xi}}))}{\sum_{i=1}^{n}T_{i}(1-\hat{e}_{i})}-\frac{\sum_{i=1}^{n}(1-T_{i})\hat{e}_{i}(Y_{i}-\check{m}(0,\bld{X}_{i};\hat{\bld{\xi}}))}{\sum_{i=1}^{n}(1-T_{i})\hat{e}_{i}},
\end{align*}
where $\check{m}(T_{i},\bld{X}_{i};\bld{\xi})$ is a outcome model.
\item BR-type estimator (proposed estimator; see Appendix \ref{appB})
$$
\hat{\tau}^{br}_{ATO}:=\frac{\sum_{i=1}^{n}\hat{e}_{i}(1-\hat{e}_{i})\tilde{m}(1,\bld{X}_{i};\hat{\bld{\xi}},\hat{\bld{\phi}})-\tilde{m}(0,\bld{X}_{i};\hat{\bld{\xi}},\hat{\bld{\phi}})}{\sum_{i=1}^{n}\hat{e}_{i}(1-\hat{e}_{i})},
$$
where
$$
\tilde{m}(T_{i},\bld{X}_{i};\bld{\xi},\bld{\phi})=\check{m}(T_{i},\bld{X}_{i};\bld{\xi})+\frac{T_{i}(1-\hat{e}_{i})}{\sum_{i=1}^{n}T_{i}(1-\hat{e}_{i})}\phi_{1}+\frac{(1-T_{i})\hat{e}_{i}}{\sum_{i=1}^{n}(1-T_{i})\hat{e}_{i}}\phi_{0},
$$
\end{itemize}
Note that $\hat{e}_{i}$ is any propensity score estimate. From the previous simulation results, we consider the methods continuously: Logistic regression, CBPS, bCART, and the proposed method. Note that the proposed method has the similar performance as the MA-based method from the result in $n=800$; in this sense, considering the proposed method only is somewhat reasonable. Estimating propensity score by each method, the estimating procedure is the same as the previous simulation (CBPS uses the covariate balancing condition (\ref{sim1})).

\begin{landscape}
\begin{figure}[h]
\begin{center}
\begin{tabular}{c}
\includegraphics[width=20cm]{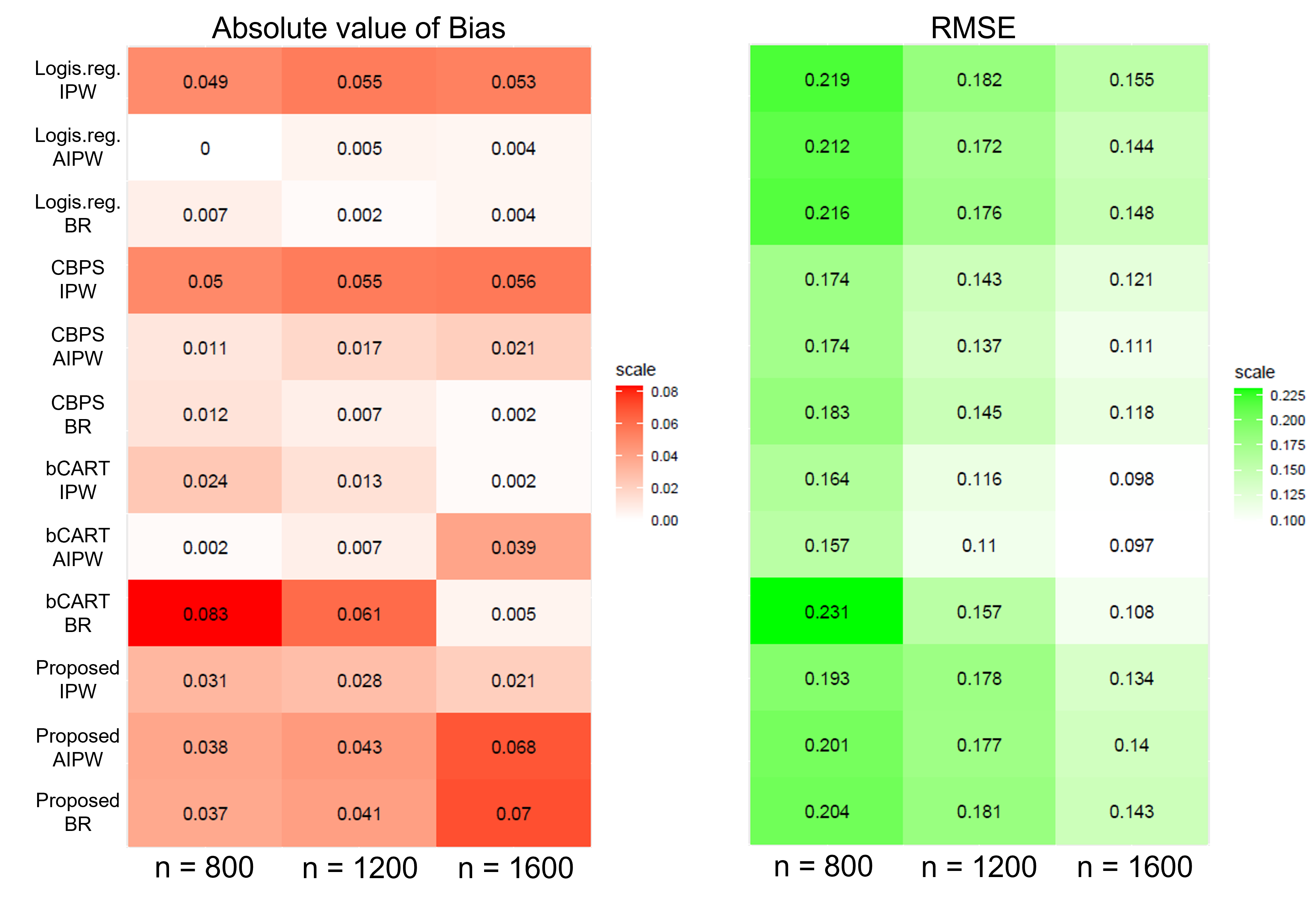}
\end{tabular}
\caption{Bias and RMSE results of each estimating methods}
\label{fig7}
\end{center}
{\footnotesize
\begin{description}
\item{\bf Note:} The simulation results of $n=1200$ are in Appendix \ref{appC}.
\end{description}
}
\end{figure}
\end{landscape}

The results are summarized in the Table \ref{tab2}, Figure \ref{fig7}, \ref{fig4}, and \ref{fig6}. As obviously, there is clear feature: BR-type estimator has good performance in the logistic regression and the CBPS estimation, and IPW estimator has good performance in the bCART and the proposed estimation (i.e., integrated method). Especially, the IPW estimator of the bCART estimation is the best in all methods; this is the same feature in the previous simulation result. Also, unstableness of the proposed method is resolved since the ATO reduses the impact of extreme propensity scores.

From here, we consider why the results of the logistic regression and the CBPS estimation (we call ``set 1''), and  the bCART and the proposed estimation (we call ``set 2'') are different. Note that the AIPW estimator is only considered since the BR estimator consistents with it (see Appendix \ref{appB}). The reason may lie in the augmentation term (see Appendix \ref{appB} also). The summary of the augmentation term is in the Table \ref{tab4}. The result of the set 1 is almost all $0$, whereas the set 2 is different from $0$. In other words, the augmentation term of the set 1 does not (needs not to) work, and the set 2 works well to adjust the bias. Therefore, the set 1 estimates the ATO without adjustment by the augmentation term even if the outcome model is misspecified. In this sense, the AIPW and BR-type estimator of the set 1 is efficient compared with the set 2.

Summarizing the above simulations, the boosted CART and CBPS with higher-order balancing condition have good properties for the ATE estimation. Also, both the CBPS with the AIPW or BR-type estimator and the boosted CART with the IPW estimator work well for the ATO estimation. The boosted CART and CBPS are useful for a variety of estimands and estimating procedures.
%\begin{landscape}
\begin{table}[htbp]
\begin{center}
\caption{Summary of estimates for the ATO}
\label{tab2}
\begin{tabular}{|c|c||c|c|c|c|}\hline
{\bf Method of}&{\bf Method of}&\multicolumn{4}{|c|}{Small sample $n=800$}\\\cline{3-6}
{\bf PS estimator}&{\bf ATO estimator}&{Mean (SD)}&{Median (Range)}&Bias&RMSE\\\hline
{\bf Logis. reg.}&IPW&0.381 (0.213)&0.393 (-0.68, 0.92)&-0.049&0.218\\\cline{2-6}
&AIPW&0.430 (0.212)&0.446 (-0.63, 0.92)&0.000&0.212\\\cline{2-6}
&BR&0.437 (0.216)&0.455 (-0.69, 0.98)&0.007&0.216\\\hline
{\bf CBPS}&IPW&0.380 (0.167)&0.381 (-0.52, 0.97)&-0.050&0.174\\\cline{2-6}
&AIPW&0.419 (0.174)&0.421 (-0.60, 0.98)&-0.011&0.174\\\cline{2-6}
&BR&0.442 (0.182)&0.443 (-0.64, 1.07)&0.012&0.183\\\hline
{\bf bCART}&IPW&0.454 (0.163)&0.446 (-0.32, 1.53)&0.024&0.164\\\cline{2-6}
&AIPW&0.432 (0.157)&0.428 (-0.33, 1.51)&0.002&0.157\\\cline{2-6}
&BR&0.513 (0.215)&0.489 (-0.26, 2.27)&0.083&0.231\\\hline
{\bf Proposed}&IPW&0.461 (0.191)&0.463 (-0.31, 1.35)&0.031&0.193\\\cline{2-6}
&AIPW&0.392 (0.197)&0.393 (-0.31, 1.43)&-0.038&0.201\\\cline{2-6}
&BR&0.393 (0.201)&0.396 (-0.31, 1.48)&-0.037&0.204\\\hline\hline
{\bf Method of}&{\bf Method of}&\multicolumn{4}{|c|}{Large sample $n=1600$}\\\cline{3-6}
{\bf PS estimator}&{\bf ATO estimator}&{Mean (SD)}&{Median (Range)}&Bias&RMSE\\\hline
{\bf Logis. reg.}&IPW&0.377 (0.145)&0.386 (-0.30, 0.75)&-0.053&0.155\\\cline{2-6}
&AIPW&0.426 (0.144)&0.435 (-0.26, 0.79)&-0.004&0.144\\\cline{2-6}
&BR&0.434 (0.148)&0.440 (-0.26, 0.80)&0.004&0.148\\\hline
{\bf CBPS}&IPW&0.374 (0.107)&0.374 (-0.02, 0.65)&-0.056&0.121\\\cline{2-6}
&AIPW&0.409 (0.109)&0.408 (0.04, 0.68)&-0.021&0.111\\\cline{2-6}
&BR&0.432 (0.118)&0.430 (0.04, 0.86)&0.002&0.119\\\hline
{\bf bCART}&IPW&0.432 (0.098)&0.434 (0.10, 0.86)&0.002&0.098\\\cline{2-6}
&AIPW&0.391 (0.089)&0.394 (0.09, 0.78)&-0.039&0.097\\\cline{2-6}
&BR&0.435 (0.108)&0.431 (0.12, 0.91)&0.005&0.108\\\hline
{\bf Proposed}&IPW&0.451 (0.132)&0.448 (0.01, 0.95)&0.021&0.134\\\cline{2-6}
&AIPW&0.362 (0.122)&0.354 (-0.04, 0.76)&-0.068&0.140\\\cline{2-6}
&BR&0.360 (0.124)&0.352 (-0.03, 0.75)&-0.070&0.143\\\hline
\end{tabular}
\end{center}
\end{table}
%\end{landscape}

\begin{table}[h]
\begin{center}
\caption{Summary of the augmentation term of the AIPW estimator}
\label{tab4}
\begin{tabular}{|c||c|}\hline
{\bf Method}&Large sample $n=1600$\\\cline{2-2}
&{Mean (SD)}\\\hline
{\bf Logis. reg.}&0.017 (0.033)\\\hline
{\bf CBPS}&0.009 (0.093)\\\hline
{\bf bCART}&0.074 (0.104)\\\hline
{\bf Proposed}&0.058 (0.072)\\\hline
\end{tabular}
\end{center}
\end{table}

\section{Discussion}
In this paper, we review the four methods, propose the novel robust two step strategy called ``integreted method''. The proposed strategies have two steps: estimating each candidate model in the first step and integrating them in the second step. Regarding the proposed integrated method, it is proved that the estimator has the $\sqrt{n}$--consistency when the true model is included in the candidate. In this sense, the proposed method has multiple robustness. Also, the KL divergence of the integrated model is smaller than each candidates when the true model is not included. In the simulation examples, the performance of these methods are compared in the situation where the estimand is the ATE and ATO, respectively. From the results of the simulation examples, the boosted CART and CBPS with higher-order balancing condition have good properties; both the estimate of the ATE and ATO has the small variance and the absolute value of bias. The boosted CART and CBPS are useful for a variety of estimands and estimating procedures.

Properties of boosted CART and CBPS are discussed in the various previous researches (e.g., Lee et al., 2010; Westreich et al., 2010; Wyss et al., 2014; Huang et al., 2022), however, to the best of our knowledge, the estimator for the ATO is not considered. Mao et al. (2019) mentioned that the impact of the proponsity score model misspecification is slight compared with the ATE. However, it is appeared that there are some impact to estimates, and the impact is different between estimating methods through our simulation result. Therefore, in spite of the estimand or estimating methods, it is appeared again that the propensity score model consideration is one of the important process in causal inference.

A different integrated method is considered in Xie et al. (2019), however, the motivation of their proposed method is slight different from our proposal. In Xie's methods, only one parametric model and one nonparametric model are integrated; in other words, taking good point of the two models. Whereas, our proposed methods are consist of parametric models only; this is one of the strong points. As mentioned in Section 2, only related to GLM procedure such as ``glm" function in R, and ``GENMOD" procedure in SAS is used. In other words, any special training is not necessary; only ``integration'' of existing knowledge for many statistics users is necessary.

In this paper, we consider robust estimation for the propensity score misspecification. On the other hand, propensity score subclassification (e.g., Wang et al., 2016; Orihara and Hamada, 2021) is another solution to ease the propensity score misspecicication problem since the propensity scores in each subclass are regarded as ``smoothed'' (Imbens and Rubin, 2015). In other words, the impact of misspecification is also smoothed. Subclassification estimator is used only for the ATE or ATT (e.g., Antonelli et al., 2018). By extending subclassification estimator, another robust ATO estimator may be derived.

\newpage
\appendix
\section{Theoretical properties of the proposed integrated procedure}
\label{appA}
\subsection{Regularity conditions}
\begin{description}
\item{{\bf C.1}} For all $k\in\{1,\dots,K\}$, $\hat{\bld{\beta}}\stackrel{P}{\to}\bld{\beta}^{*},$ where $\bld{\beta}^{*}$ is the ``true" parameter value (see White, 1982).
\item{{\bf C.2}} For all $\theta\in\mathbb{R}$, $\ddot{b}(\theta)>0$, where $\ddot{b}$ is the second differentiables of $b$.
\item{{\bf C.3}} $^\exists\bld{x}\in\mathbb{R}^{p}\ s.t.\ \theta_{k}(\bld{x};\bld{\beta}_{k}^{*})\neq \theta_{k'}(\bld{x};\bld{\beta}_{k'}^{*})$ for all combinations of $k\neq k'$.
\item{{\bf C.4}}  $^\forall \delta_{n}\to0$,
$$
\sup_{\bld{\gamma}:||\bld{\gamma}-\bld{\gamma}^{0}||<\delta_{n}}\frac{|R(\bld{\gamma})|}{(1+||\sqrt{n}(\bld{\gamma}-\bld{\gamma}^{0})||)^2}=o_{p}(1),
$$
where $\bld{\gamma}^{0}=(1,0,\dots,0)^{\top}$, and $R(\bld{\gamma})$ is $3^{rd}$ remainder term of the taylor expansion of the likelihood function of the integrated model around $\bld{\gamma}^{0}$.
\end{description}

\subsection{Constructing a robust estimating method under GLM}
Let $n$ be the sample size, and assume that $i = 1,\ 2,\dots, n$ are i.i.d. samples. $t\in\mathcal{T}\subset\mathbb{R}$ and $\bld{X}\in\mathbb{R}^{p}$ denote an outcome and a vector of covariates, respectively. We assume that an outcome has a distribution in an exponential family (McCullagh and Nelder, 2019):
\begin{align}
\label{eq1}
f(t;\theta,\phi)=\exp\left\{\frac{t\theta-b(\theta)}{a(\phi)}+c(t,\phi)\right\},
\end{align}
where $(\theta,\phi)$ are parameters; in particular, we are interested in the parameter $\theta$. For a normal distribution (i.e., for the generalized propensity score), each function and parameter in (\ref{eq1}) become
$$
\theta=\mu\ (mean),\ \ a(\phi)=\sigma^2\ (variance),\ \ b(\theta)=\frac{\theta^2}{2},\ \ c(y,\phi)=-\frac{y^2}{2\sigma^2}-\frac{1}{2}\log(2\pi\sigma^2).
$$
For a Bernoulli distribution (i.e., for the propensity score), each function and parameter in (\ref{eq1}) become
$$
\theta=\log\left(\frac{p}{1-p}\right)\ (p\ is\ a\ binomial\ probability),\ \ a(\phi)=1,\ \ b(\theta)=\log(1+e^{\theta}),\ \ c(y,\phi)=0.
$$

To construct a GLM, we consider a relationship between an expectation of an outcome and its model structure $\varphi(\bld{x};\bld{\beta})$. Specifically, $\varphi(\bld{x};\bld{\beta})=g(p)$, where a monotonic function $g$ is called ``link function". For a Bernoulli distribution, when we select a link function as $g=\log\ (logit\ link)$ and assume a linear relationship between $\bld{x}$ and $\bld{\beta}$ ($\varphi(\bld{x};\bld{\beta})=\bld{x}^{\top}\bld{\beta}$),
$$
p=\frac{\exp\left\{\bld{x}^{\top}\bld{\beta}\right\}}{1+\exp\left\{\bld{x}^{\top}\bld{\beta}\right\}}\ \ \ (logistic\ regression\ model).
$$
Whereas, when we select a link function as $g=\Phi^{-1}\ (a\ normal\ distribution\ function;\ probit\ link)$ and assume a linear relationship between $\bld{x}$ and $\bld{\beta}$,
$$
p=\Phi\left(\bld{x}^{\top}\bld{\beta}\right)\ \ \ (probit\ model).
$$
Note that $\theta$ becomes a function of $\bld{x}$ and $\bld{\beta}$ clearly from the above example: $\theta=\theta(\bld{x};\bld{\beta})$. Hereinafter, the model (\ref{eq1}) will be considered for a while.

From here, we introduce a proposed procedure that many working models are combined into a one integrated model. At first, we consider working models; $k\, (k=1,2,\dots,K)$ denotes each model:
$$
f(t|\bld{x};\bld{\beta}_{k})=\exp\left\{\frac{t\theta_{k}(\bld{x};\bld{\beta}_{k})-b(\theta_{k}(\bld{x};\bld{\beta}_{k}))}{a(\phi)}+c(t,\phi)\right\}.
$$
Note that we assume that $\phi$ is known and needs not to be estimated, and the functions $a$, $b$, and $c$ are common to each model. For a normal distribution, these assumptions imply that variance parameters are known and common to each model. To estimate parameter $\bld{\beta}_{k}$, we consider a KL divergence $D_{k}(\bld{\beta}_{k})$:
\begin{align}
\label{eq3}
D_{k}(\bld{\beta}_{k})={\rm E}\left[\log g(T|\bld{X})\right]-{\rm E}\left[\log f(T|\bld{X};\bld{\beta}_{k})\right],
\end{align}
where $g(t|\bld{x})$ is the unknown true probability distribution. Whereas we would like to estimate the parameters so that $D_{k}(\bld{\beta}_{k})$ is small, we cannot handle (\ref{eq3}) directly. Therefore, we consider an estimator of the second term of (\ref{eq3}) to estimate parameter $\bld{\beta}_{k}$: 
$$
\ell(\bld{\beta}_{k})=\frac{1}{n}\sum_{i=1}^{n}\log f(t_{i}|\bld{x}_{i};\bld{\beta}_{k})\propto \frac{1}{n}\sum_{i=1}^{n}\left[t_{i}\theta_{k}(\bld{x}_{i};\bld{\beta}_{k})-b(\theta_{k}(\bld{x}_{i};\bld{\beta}_{k}))\right],
$$
where $\ell$ is a log-likelihood function. Therefore, a score function $S(\bld{\beta}_{k})$ becomes
$$
S(\bld{\beta}_{k})=\frac{\partial}{\partial \bld{\beta}_{k}}\ell(\bld{\beta}_{k})=\frac{1}{n}\sum_{i=1}^{n}\left[t_{i}-\dot{b}(\theta_{k}(\bld{x}_{i};\bld{\beta}_{k}))\right]\frac{\partial}{\partial \bld{\beta}_{k}}\theta_{k}(\bld{x}_{i};\bld{\beta}_{k}),
$$
where $\dot{b}$ is the first differentiable of $b$. The solution of $S(\bld{\beta}_{k})=\bld{0}$ becomes a maximum likelihood estimator (MLE) $\hat{\bld{\beta}}_{k}$.

Next, by using the MLEs $\hat{\bld{\beta}}_{k}$, we consider an integrated model:
$$
f(t|\bld{x};\bld{\gamma},\hat{\bld{\beta}})=\exp\left\{\frac{t\sum_{k=1}^{K}\gamma_{k}\theta_{k}(\bld{x};\hat{\bld{\beta}}_{k})-b\left(\sum_{k=1}^{K}\gamma_{k}\theta_{k}(\bld{x};\hat{\bld{\beta}}_{k})\right)}{a(\phi)}+c(t,\phi)\right\},
$$
where $\gamma_{k}\in (0,1)$. The same discussion as above, a KL divergence, a log-likelihood function, and a score function become
\begin{align}
\label{eq4}
D(\bld{\gamma},\bld{\beta})={\rm E}\left[\log g(T|\bld{X})\right]-{\rm E}\left[\log f(T|\bld{X};\bld{\gamma},\bld{\beta})\right],
\end{align}
$$
\ell(\bld{\gamma},\hat{\bld{\beta}})\propto\frac{1}{n}\sum_{i=1}^{n}\left[t_{i}\sum_{k=1}^{K}\gamma_{k}\theta_{k}(\bld{x}_{i};\hat{\bld{\beta}}_{k})-b\left(\sum_{k=1}^{K}\gamma_{k}\theta_{k}(\bld{x}_{i};\hat{\bld{\beta}}_{k})\right)\right],
$$
\begin{align}
\label{eq7}
S(\bld{\gamma},\hat{\bld{\beta}})=\frac{\partial}{\partial \bld{\gamma}}\ell(\bld{\gamma},\hat{\bld{\beta}})=\frac{1}{n}\sum_{i=1}^{n}\left[t_{i}-\dot{b}\left(\sum_{k'=1}^{K}\gamma_{k'}\theta_{k'}(\bld{x}_{i};\hat{\bld{\beta}}_{k'})\right)\right]\left(
\begin{array}{c}
\theta_{1}(\bld{x}_{i};\hat{\bld{\beta}}_{1})\\
\vdots\\
\theta_{K}(\bld{x}_{i};\hat{\bld{\beta}}_{K})
\end{array}
\right),
\end{align}
The solution of $S(\bld{\gamma},\hat{\bld{\beta}})=\bld{0}$ becomes a MLE $\hat{\bld{\gamma}}$, and the integrated model $f(t|\bld{x};\hat{\bld{\gamma}},\hat{\bld{\beta}})$ is used for a subsequent inference.

In the next step, we confirm properties of our proposed estimating procedure. At first, we consider the situation where the true model is included in the candidate models ($k=1,2,\dots,K$). Concretely, we assume that $k=1$ is the true model. Under this setting, the following theorem is proved:
\begin{theo}{$\phantom{}$}\\
Under the regularity conditions from C.1 to C.3,
\begin{align}
\label{eq5}
\hat{\gamma}_{1}\stackrel{P}{\to}1\ \ and\ \ \hat{\gamma}_{k}\stackrel{P}{\to}0,\ k=2,\dots,K.
\end{align}
Therefore, from the continuous mapping theorem,
\begin{align*}
f(t|\bld{x};\hat{\bld{\gamma}},\hat{\bld{\beta}})&=\exp\left\{\frac{t\sum_{k=1}^{K}\hat{\gamma}_{k}\theta_{k}(\bld{x};\hat{\bld{\beta}}_{k})-b\left(\sum_{k=1}^{K}\hat{\gamma}_{k}\theta_{k}(\bld{x};\hat{\bld{\beta}}_{k})\right)}{a(\phi)}+c(t,\phi)\right\}\\
&\stackrel{P}{\to}\exp\left\{\frac{t\theta_{1}(\bld{x};\bld{\beta}_{1}^{0})-b\left(\bld{\beta}_{1}^{0}\right)}{a(\phi)}+c(t,\phi)\right\}\\
&=f(t|\bld{x};\bld{\beta}_{1}^{0}),
\end{align*}
where the superscript ``0" of parameters means the true value of parameters.
\end{theo}\noindent
Proof is in Appendix \ref{appA3}. From the Theorem 1, the integrated model is consistent with the true model when the true model is included in the candidate models, and $n\to\infty$. Also, assuming additional regularity condition, the $\sqrt{n}$-consistency is also proved (see Andrews, 1999).
\begin{theo}{$\phantom{}$}\\
Under the regularity conditions from C.1 to C.4,
$$
\sqrt{n}\left(\hat{\bld{\gamma}}-\bld{\gamma}^{0}\right)=O_{p}(1),
$$
where $\bld{\gamma}^{0}=(1,0,\dots,0)^{\top}$.
\end{theo}\noindent
Proof is straightforward from Theorem 1 and the result of Andrews (1999).

Next, we consider the situation where the true model is not included in the candidate models; the assumption of Theorem 1 and 2 is not hold. Even in this situation, the following theorem is hold:
\begin{theo}{$\phantom{}$}\\
Additionally, assuming that $\sum_{k=1}^{K}\gamma_{k}=1$. Under the regularity condition C.2, the following inequality is hold for $\forall\bld{\gamma}$:
\begin{align}
\label{eq6}
D(\bld{\gamma},\bld{\beta}^{*})\leq\sum_{k=1}^{K}\gamma_{k}D_{k}(\bld{\beta}_{k}^{*})
\end{align}
\end{theo}\noindent
Proof is in Appendix \ref{appA3}. The Theorem 3 means that the integrated model becomes better than each candidate model in the sense of the true KL divergence even if the true model is not included in the candidate models. In this sense, the integrated model is better option than any model selection properties when we are not interested in the selection of the ``valid" model. Note that it is necessary to choose no hyper parameters in our proposed method; this is a different point from any machine learning procedures.

Our proposed method can be applied usual propensity score based procedures. For instance, IPW estimator, matching estimator, and subclassification estimator for the ATE. However, one of the important notations is that the weight estimators for each candidates models ($\hat{\bld{\gamma}}$) only have the $\sqrt{n}$-consistency; not the asymptotic normality. Theoretically, this difference is important since the propensity score based procedures have different property compared with using only one propensity score estimator. Applicationally, constructing confidence intervals by bootstrap method is necessary. Since our proposed method is based on GLM, the application for the prognostic score (Hansen, 2008) is also considered. As described in Antonelli et al. (2018), the doubly robust matching estimator, in the sense that the estimator becomes the consistency either the propensity score model or the prognostic score model is correctly specified, is proposed. By applying our proposed method, the multply robust matching estimator can be constructed.

\subsection{Proofs}
\label{appA3}
\subsubsection{Proof of Theorem 1}
From C.1 and the continuous mapping theorem, (\ref{eq7}) becomes
$$
S(\bld{\gamma},\hat{\bld{\beta}})=S(\bld{\gamma},\bld{\beta}^{*})+o_{p}(1).
$$
Also, from (\ref{eq7}),
\begin{align}
\label{eq8}
\frac{\partial^2}{\partial \bld{\gamma}^{\otimes 2}}\ell(\bld{\gamma},\bld{\beta}^{*})=\frac{\partial}{\partial \bld{\gamma}}S^{\top}(\bld{\gamma},\bld{\beta}^{*})=-\frac{1}{n}\sum_{i=1}^{n}\ddot{b}\left(\sum_{k'=1}^{K}\gamma_{k'}\theta_{k'}(\bld{x}_{i};\bld{\beta}^{*}_{k'})\right)\left(
\begin{array}{c}
\theta_{1}(\bld{x}_{i};\bld{\beta}^{0}_{1})\\
\theta_{2}(\bld{x}_{i};\bld{\beta}^{*}_{2})\\
\vdots\\
\theta_{K}(\bld{x}_{i};\bld{\beta}^{*}_{K})
\end{array}
\right)^{\otimes 2}.
\end{align}
Therefore, from C.2, (\ref{eq8}) becomes the negative definite:
\begin{align}
\label{eq9}
\frac{\partial^2}{\partial \bld{\gamma}^{\otimes 2}}\ell(\bld{\gamma},\bld{\beta}^{*})<O.
\end{align}
From (\ref{eq9}), (\ref{eq7}) has the unique solution $\hat{\bld{\gamma}}\stackrel{P}{\to}\bld{\gamma}^{0}$ satisfying
\begin{align}
\label{eq10}
{\rm E}\left[\left[T-\dot{b}\left(\sum_{k'=1}^{K}\gamma_{k'}^{0}\theta_{k'}(\bld{X};\bld{\beta}^{*}_{k'})\right)\right]\left(
\begin{array}{c}
\theta_{1}(\bld{X};\bld{\beta}^{0}_{1})\\
\theta_{2}(\bld{X};\bld{\beta}^{*}_{2})\\
\vdots\\
\theta_{K}(\bld{X};\bld{\beta}^{*}_{K})
\end{array}
\right)\right]=\bld{0}
\end{align}

Next, the concrete values of $\bld{\gamma}^{0}$ is confirmed. The expectation regarding $T|\bld{x}$ of (\ref{eq10}) is
\begin{align}
\label{eq11}
{\rm E}\left[\left.T-\dot{b}\left(\sum_{k'=1}^{K}\gamma_{k'}^{0}\theta_{k'}(\bld{x};\bld{\beta}^{*}_{k'})\right)\right|\bld{x}\right]
\end{align}
From the property of GLM (see McCullagh and Nelder, 2019),
$$
{\rm E}\left[\left.T\right|\bld{x}\right]=\dot{b}(\theta_{1}(\bld{x};\bld{\beta}^{0}_{1})).
$$
Therefore, (\ref{eq11}) becomes
$$
\dot{b}(\theta_{1}(\bld{x};\bld{\beta}^{0}_{1}))-\dot{b}\left(\sum_{k'=1}^{K}\gamma_{k'}^{0}\theta_{k'}(\bld{x};\bld{\beta}^{*}_{k'})\right).
$$
From the above, (\ref{eq10}) becomes
\begin{align}
\label{eq12}
{\rm E}\left[\left[\dot{b}(\theta_{1}(\bld{X};\bld{\beta}^{0}_{1}))-\dot{b}\left(\sum_{k'=1}^{K}\gamma_{k'}^{0}\theta_{k'}(\bld{X};\bld{\beta}^{*}_{k'})\right)\right]\left(
\begin{array}{c}
\theta_{1}(\bld{X};\bld{\beta}^{0}_{1})\\
\theta_{2}(\bld{X};\bld{\beta}^{*}_{2})\\
\vdots\\
\theta_{K}(\bld{X};\bld{\beta}^{*}_{K})
\end{array}
\right)\right]=\bld{0}.
\end{align}
From C.3, there exists the unique value of $\bld{\gamma}$ such that
$$
\gamma^{0}_{1}=1\ \ and\ \ \gamma^{0}_{k}=0,\ k=2,\dots,K.
$$
Therefore, (\ref{eq5}) is obtained.

\subsubsection{Proof of Theorem 3}
It is sufficient to show that
$$
-{\rm E}\left[\log f(T|\bld{X};\bld{\gamma},\bld{\beta}^{*})\right]+\sum_{k=1}^{K}\gamma_{k}{\rm E}\left[\log f(T|\bld{X};\bld{\beta}^{*}_{k})\right]\leq0,
$$
i.e.,
\begin{align}
\label{eq20}
-\int \left(t\sum_{k=1}^{K}\gamma_{k}\theta_{k}(\bld{x};\bld{\beta}^{*}_{k})-b\left(\sum_{k=1}^{K}\gamma_{k}\theta_{k}(\bld{x};\bld{\beta}^{*}_{k})\right)\right)f(y,\bld{x})dyd\bld{x}&\nonumber\\
&\hspace{-7.5cm}+\sum_{k=1}^{K}\gamma_{k}\left[\int\left(t\theta_{k}(\bld{x};\bld{\beta}^{*}_{k})-b(\theta_{k}(\bld{x};\bld{\beta}^{*}_{k}))\right)f(t,\bld{x})dyd\bld{x}\right]\nonumber\\
&\hspace{-10.5cm}=\int b\left(\sum_{k=1}^{K}\gamma_{k}\theta_{k}(\bld{x};\bld{\beta}^{*}_{k})\right)f(\bld{x})d\bld{x}-\sum_{k=1}^{K}\gamma_{k}\int b(\theta_{k}(\bld{x};\bld{\beta}^{*}_{k}))f(\bld{x})d\bld{x}\nonumber\\
&\hspace{-10.5cm}=\int \left[b\left(\sum_{k=1}^{K}\gamma_{k}\theta_{k}(\bld{x};\bld{\beta}^{*}_{k})\right)-\sum_{k=1}^{K}\gamma_{k}\int b(\theta_{k}(\bld{x};\bld{\beta}^{*}_{k}))\right]f(\bld{x})d\bld{x}\leq 0.
\end{align}
From C.2, by using the property of convex functions,
$$
b\left(\sum_{k=1}^{K}\gamma_{k}\theta_{k}\right)\leq \sum_{k=1}^{K}\gamma_{k}b(\theta_{k})
$$
for all $\theta_{k}$ and $\bld{\gamma}$. Therefore, (\ref{eq20}) is hold; (\ref{eq6}) is obtained.

\section{Construction of BR-type estimator for ATO}
\label{appB}
To estimate the ATO, we introduce the previous two estimators: the IPW-type and the AIPW-type estimators.
\begin{itemize}
\item IPW-type estimator (Li et al., 2018)
$$
\hat{\tau}^{ipw}_{ATO}:=\frac{\sum_{i=1}^{n}T_{i}(1-e_{i})Y_{i}}{\sum_{i=1}^{n}T_{i}(1-e_{i})}-\frac{\sum_{i=1}^{n}(1-T_{i})e_{i}Y_{i}}{\sum_{i=1}^{n}(1-T_{i})e_{i}}.
$$
\item AIPW-type estimator (Mao et al., 2019)
\begin{align}
\label{seq1}
\hat{\tau}^{aug}_{ATO}&:=\frac{\sum_{i=1}^{n}e_{i}(1-e_{i})(m_{1i}-m_{0i})}{\sum_{i=1}^{n}e_{i}(1-e_{i})}+\frac{\sum_{i=1}^{n}T_{i}(1-e_{i})(Y_{i}-m_{1i})}{\sum_{i=1}^{n}T_{i}(1-e_{i})}-\frac{\sum_{i=1}^{n}(1-T_{i})e_{i}(Y_{i}-m_{0i})}{\sum_{i=1}^{n}(1-T_{i})e_{i}}\nonumber\\
\end{align}
where $m_{ti}={\rm E}[Y_{t}|\bld{X}=\bld{x}_{i}],\, t\in\{0,1\}$ is the true outcome model.
\end{itemize}
Focusing on the first term of (\ref{seq1}), the denominator becomes
\begin{align*}
\frac{1}{n}\sum_{i=1}^{n}e_{i}(1-e_{i})(m_{1i}-m_{0i})&\stackrel{P}{\to}{\rm E}\left[e(\bld{X})(1-e(\bld{X}))(m_{1}(\bld{X})-m_{0}(\bld{X}))\right]\\
&={\rm E}\left[e(\bld{X})(1-e(\bld{X})){\rm E}[Y_{1}-Y_{0}|\bld{X}]\right].
\end{align*}
Therefore, 
$$
\frac{\sum_{i=1}^{n}e_{i}(1-e_{i})(m_{1i}-m_{0i})}{\sum_{i=1}^{n}e_{i}(1-e_{i})}\stackrel{P}{\to}\tau_{ATO}.
$$
From the above, AIPW-type estimator (\ref{seq1}) can be considered as
$$
\hat{\tau}^{aug}_{ATO}=``Main\ term" + ``Augmentation\ term".
$$
The augmentation term is involved in the efficiency improvement of IPW-type estimator.

From here, we consider the boundedness; when considering the property, we assume $\mathcal{Y}=\{0,1\}^{\otimes 2}$. Actually, the AIPW-estimator (\ref{seq1}) does not have the property. Obviously, the first term falls within $(-1,1)$. Whereas, the augmentation terms cause the break-down of the boundedness. To confirm it, we consider an extreme example: almost all subjects are $T_{i}=1$, $e_{i}\approx 0$, $Y_{i}=0$, and $m_{1i}\approx 1$. Then, the second and third terms become
$$
\frac{\sum_{i=1}^{n}T_{i}(1-e_{i})(Y_{i}-m_{1i})}{\sum_{i=1}^{n}T_{i}(1-e_{i})}\approx -1,\ \ \frac{\sum_{i=1}^{n}(1-T_{i})e_{i}(Y_{i}-m_{0i})}{\sum_{i=1}^{n}(1-T_{i})e_{i}}\approx 0.
$$
This is very rare or unusual situation, however, the AIPW-type estimator may fail the boundedness.

To overcome the problem, the novel BR-type estimator is derived. The main flow is the same as the Bang and Robins' estimator for the ATE; more precisely, we consider the outcome regression based doubly robust estimator. To derive the estimator, we consider the following estimating equation:
\begin{align}
\label{seq2}
\sum_{i=1}^{n}\left(
\begin{array}{c}
\frac{\partial}{\partial\bld{\xi}}\tilde{m}(T_{i},\bld{X}_{i};\bld{\xi},\bld{\phi})\\
\frac{T_{i}(1-e(\bld{X}_{i};\hat{\bld{\beta}}))}{\sum_{i=1}^{n}T_{i}(1-e(\bld{X}_{i};\hat{\bld{\beta}}))}\\
\frac{(1-T_{i})e(\bld{X}_{i};\hat{\bld{\beta}})}{\sum_{i=1}^{n}(1-T_{i})e(\bld{X}_{i};\hat{\bld{\beta}})}
\end{array}
\right)(Y_{i}-\tilde{m}(T_{i},\bld{X}_{i};\bld{\xi},\bld{\phi}))=\bld{0},
\end{align}
where
$$
\tilde{m}(T_{i},\bld{X}_{i};\bld{\xi},\bld{\phi})=\check{m}(T_{i},\bld{X}_{i};\bld{\xi})+\frac{T_{i}(1-e(\bld{X}_{i};\hat{\bld{\beta}}))}{\sum_{i=1}^{n}T_{i}(1-e(\bld{X}_{i};\hat{\bld{\beta}}))}\phi_{1}+\frac{(1-T_{i})e(\bld{X}_{i};\hat{\bld{\beta}})}{\sum_{i=1}^{n}(1-T_{i})e(\bld{X}_{i};\hat{\bld{\beta}})}\phi_{0},
$$
$\check{m}(T_{i},\bld{X}_{i};\bld{\xi})$ is an outcome model, and $e(\bld{X}_{i};\hat{\bld{\beta}})$ are estimated propensity scores. In other words, estimators $\hat{\bld{\xi}}$ and $\hat{\bld{\phi}}$ are the solutions of (\ref{seq2}). To estimate the ATO, constructing the following estimator:
\begin{itemize}
\item BR-type estimator (proposed estimator)
\begin{align}
\label{seq3}
\hat{\tau}^{br}_{ATO}:=\frac{\sum_{i=1}^{n}e_{i}(1-e_{i})\tilde{m}(1,\bld{X}_{i};\hat{\bld{\xi}},\hat{\bld{\phi}})-\tilde{m}(0,\bld{X}_{i};\hat{\bld{\xi}},\hat{\bld{\phi}})}{\sum_{i=1}^{n}e_{i}(1-e_{i})}
\end{align}
\end{itemize}
The proposed BR-type estimator is known as a ``plug-in" type estimator since treatment values $t=0$ and $1$ are plugged into the estimated outcome models $\tilde{m}(t,\bld{X}_{i};\hat{\bld{\xi}},\hat{\bld{\phi}})$. This type estimator is easy to calculate, and will be used not only in academia but also industry (see FDA guidance, 2019). Note that $\hat{\tau}^{br}_{ATO}$ has the boundedness obviously.

From here, we consider the properties of the proposed BR-type estimator.
\begin{theo}{$\phantom{}$}\\
\label{theo1}
If propensity score models $e(\bld{X}_{i};\bld{\beta})$ are correctly specified, then $\hat{\tau}^{br}_{ATO}\stackrel{P}{\to}\tau_{ATO}$. Additionally, outcome models $\check{m}(T_{i},\bld{X}_{i};\bld{\xi})$ are correctly specified, then $\hat{\tau}^{br}_{ATO}$ has the semiparametric efficiency.
\end{theo}\noindent
The statement of {\bf Theorem \ref{theo1}.} is different from the ordinary double robustness: the propensity score models need to be specified. Since the proof is the same flow as p.963-964 of Bang and Robins (2005), we do not provide it in this paper.

\section{Simulation results}
\label{appC}
\subsection{Application to the ATE}
\begin{comment}
\begin{table}[h]
\begin{center}
\caption{Summary of estimates for the ATE}
\begin{tabular}{|c||c|c|c|c|}\hline
{\bf Method}&\multicolumn{4}{|c|}{Small sample $n=1200$}\\\cline{2-5}
&{Mean (SD)}&{Median (Range)}&Bias&RMSE\\\hline
{\bf Logis. reg.}&0.368 (0.225)&0.38 (-1.34, 0.98)&-0.132&0.261\\\hline
{\bf CBPS}&0.387 (0.156)&0.402 (-0.32, 0.92)&-0.113&0.192\\\hline
{\bf bCART}&0.616 (0.162)&0.619 (0.06, 2.27)&0.116&0.200\\\hline
{\bf Liu et al.}&0.382 (0.179)&0.386 (-0.04, 0.89)&-0.118&0.214\\\hline
{\bf Proposed}&0.522 (0.707)&0.561 (-12.84, 4.23)&0.022&0.707\\\hline
{\bf MA-based}&0.065 (1.125)&0.065 (-0.73, 0.86)&-0.435&1.206\\\hline
\end{tabular}
\end{center}
{\footnotesize
\begin{description}
\item{\bf Note 1:} Some propoensity score estimaties of Liu's method become ``NA''. The number that at reast one propensity score estimate becomes ``NA'' is 327 (65.4\%) in $n=800$, and 386 (77.2\%) in $n=1600$, respectively. The statistics summarized in Table \ref{tab1} are calculated from the time there is no ``NA'' values (i.e., $500-327=173$ in $n=800$, and $500-386=114$ in $n=1600$).
\item{\bf Note 2:} The weights $w$ of MA-based method become $0$ in $n=1600$. This is because $\exp\left\{-\frac{IC_{k}}{2}\right\}$ becomes very small value, and is handled as $0$ in R. Therefore, the estimates of MA-based cannot be calculated.
\end{description}
}
\end{table}
\end{comment}

\begin{landscape}
\begin{figure}[h]
\begin{center}
\begin{tabular}{c}
\includegraphics[width=20cm]{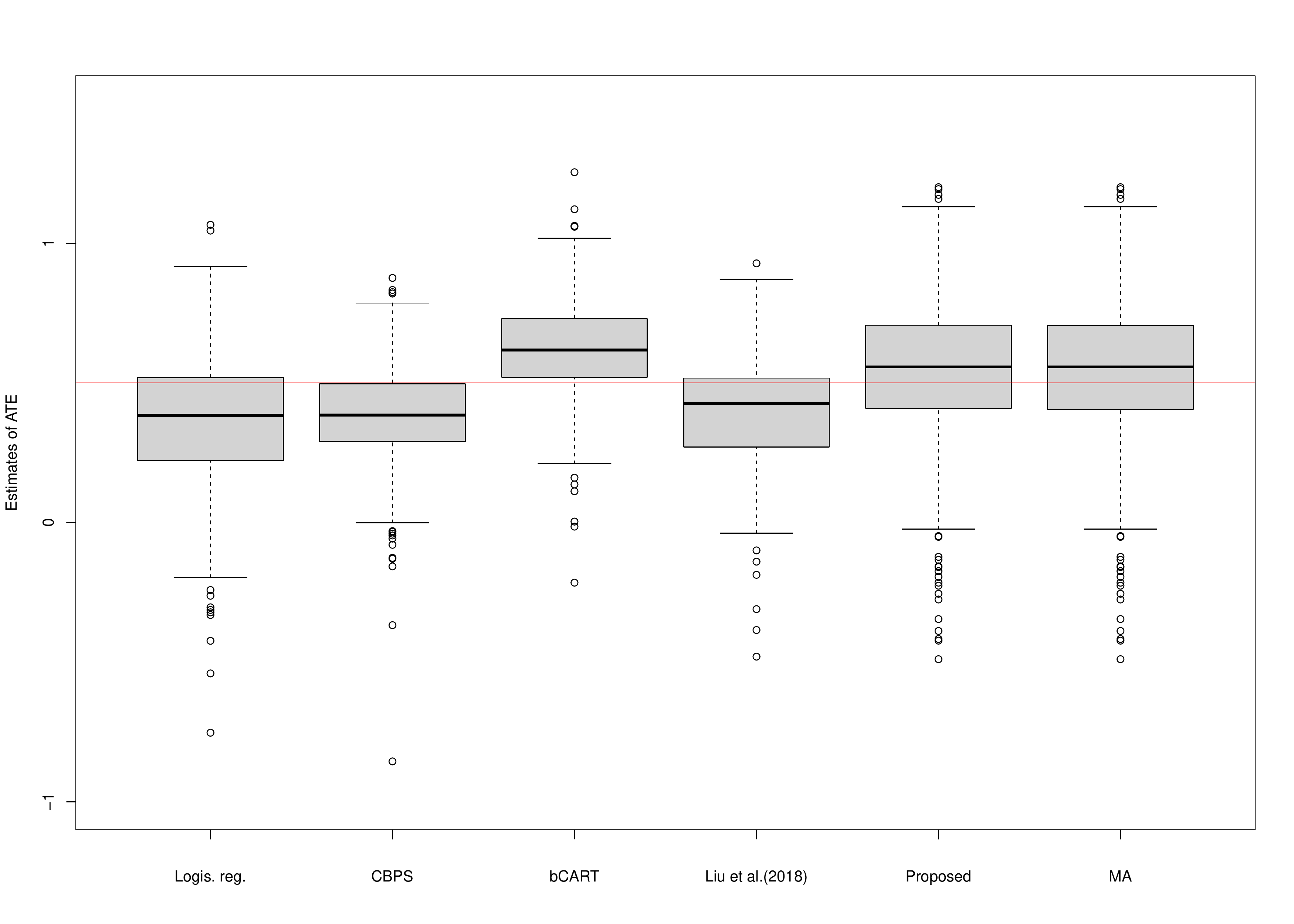}
\end{tabular}
\caption{Simulation results ($n=800$)}
\label{fig1}
\end{center}
\end{figure}

\begin{comment}
\begin{figure}[h]
\begin{center}
\begin{tabular}{c}
\includegraphics[width=20cm]{kk_plot_n1200_K500.pdf}
\end{tabular}
\caption{Simulation results ($n=1200$)}
\label{fig2}
\end{center}
\end{figure}
\end{comment}

\begin{figure}[h]
\begin{center}
\begin{tabular}{c}
\includegraphics[width=20cm]{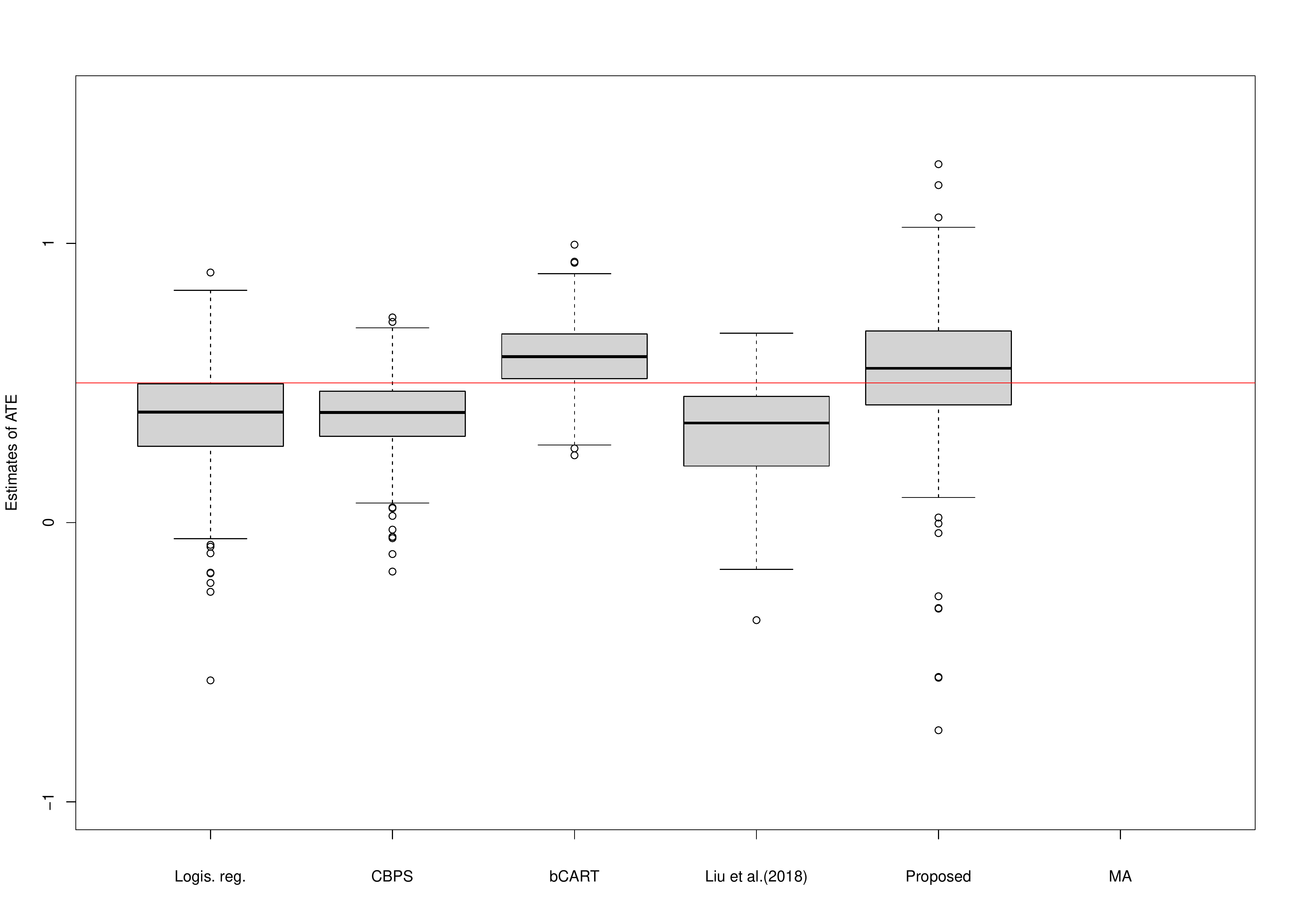}
\end{tabular}
\caption{Simulation results ($n=1600$)}
\label{fig3}
\end{center}
\end{figure}

\begin{figure}[h]
\begin{center}
\begin{tabular}{c}
\includegraphics[width=24cm]{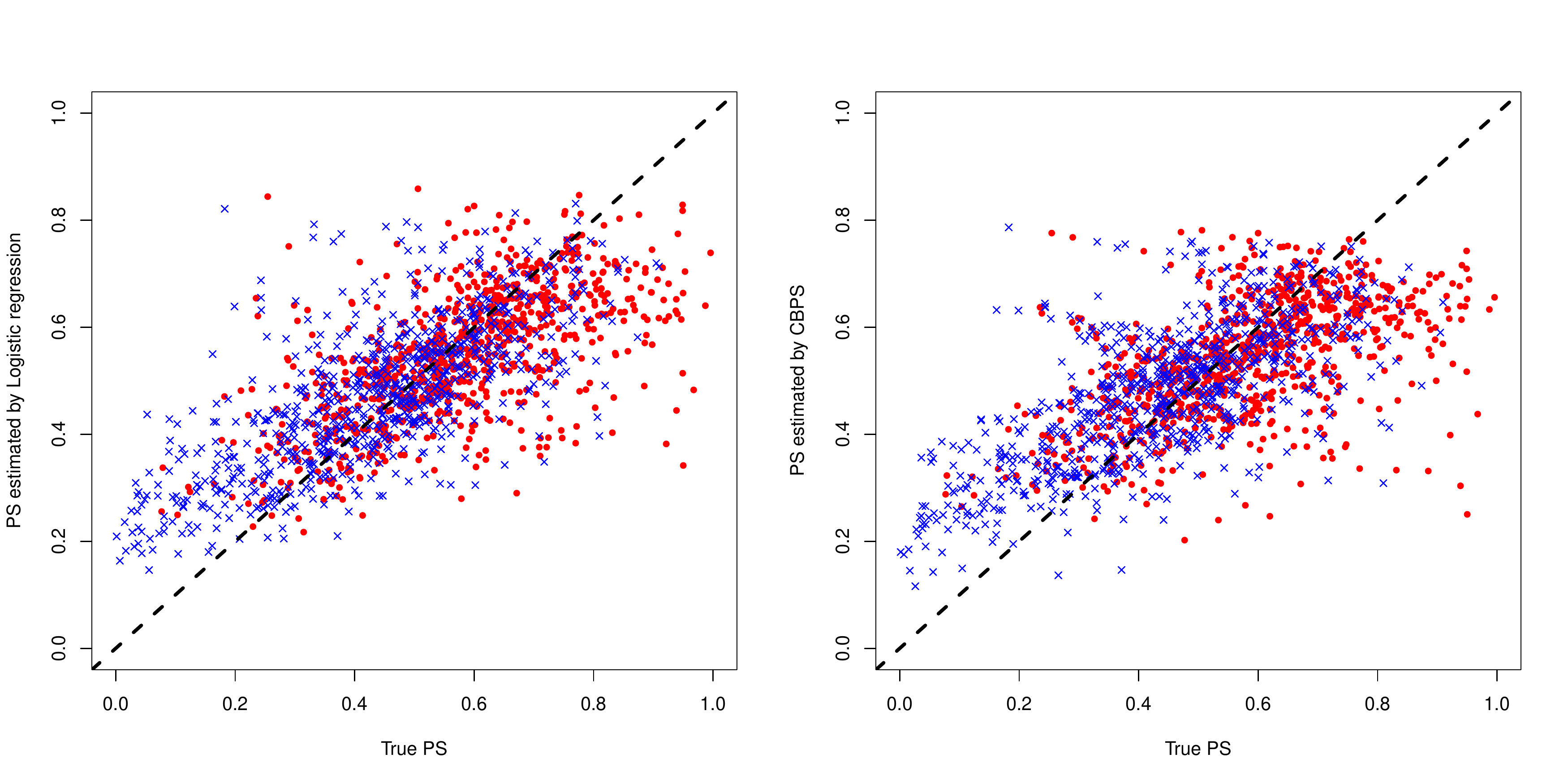}
\end{tabular}
\caption{Scatter plots of each estimating methods in an iteration (in the appendix) (1/2)}
\end{center}
{\footnotesize
\begin{description}
\item{\bf Note:} Red circle is the treatment group ($T=1$) and the blue cross is the control group ($T=0$).
\end{description}
}
\end{figure}

\begin{figure}[h]
\begin{center}
\begin{tabular}{c}
\includegraphics[width=24cm]{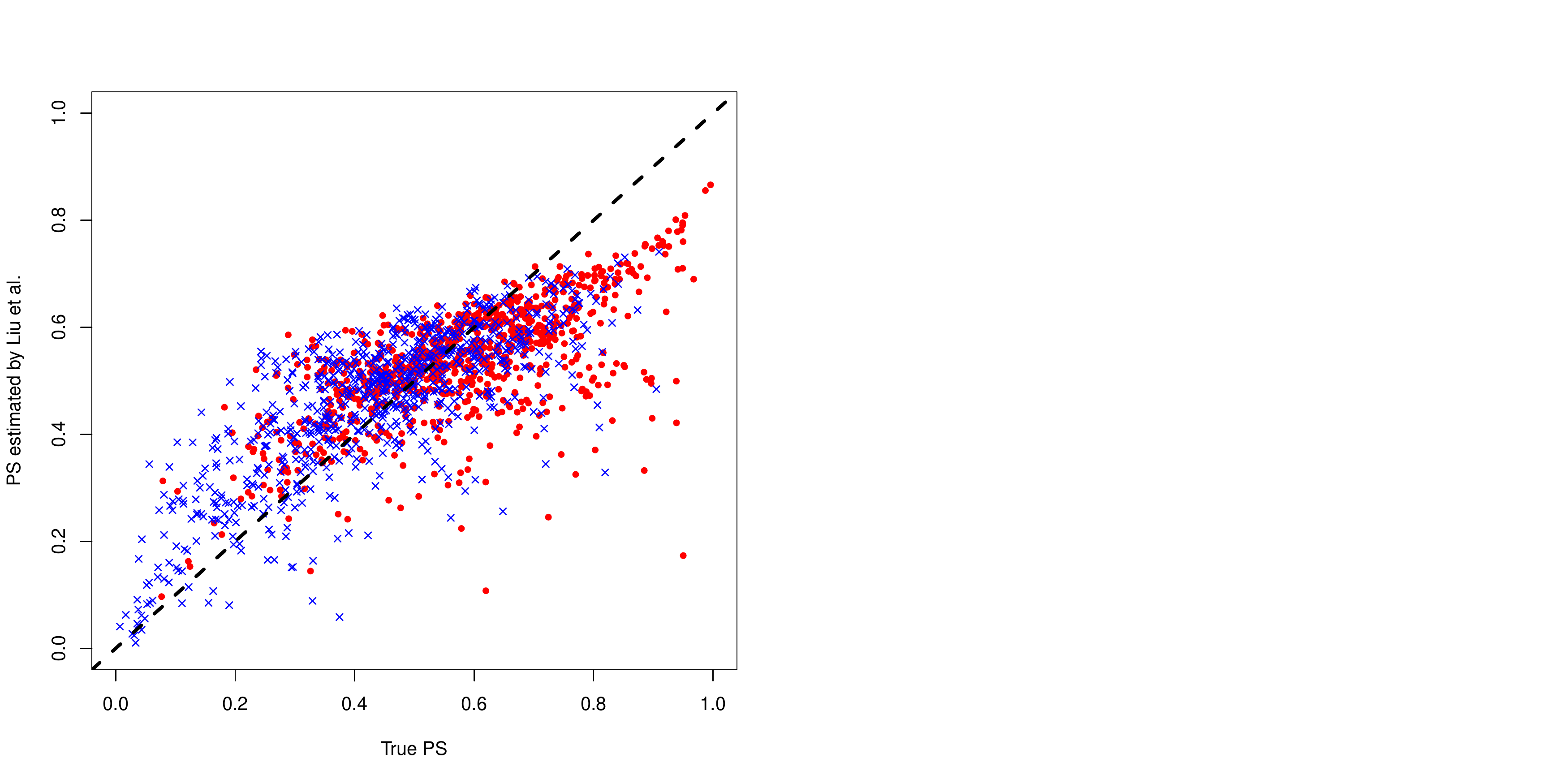}
\end{tabular}
\caption{Scatter plots of each estimating methods in an iteration (in the appendix) (2/2)}
\end{center}
{\footnotesize
\begin{description}
\item{\bf Note:} Red circle is the treatment group ($T=1$) and the blue cross is the control group ($T=0$).
\end{description}
}
\end{figure}

\end{landscape}
\subsection{Application to the ATO}
The results of $n=1200$ are summarized in the Table \ref{tab3}, Figure \ref{fig5}.
\begin{table}[htbp]
\begin{center}
\caption{Summary of estimates for the ATO}
\label{tab3}
\begin{tabular}{|c|c||c|c|c|c|}\hline
{\bf Method of}&{\bf Method of}&\multicolumn{4}{|c|}{Small sample $n=1200$}\\\cline{3-6}
{\bf PS estimator}&{\bf ATO estimator}&{Mean (SD)}&{Median (Range)}&Bias&RMSE\\\hline
{\bf Logis. reg.}&IPW&0.375 (0.174)&0.379 (-0.30, 0.93)&-0.055&0.182\\\cline{2-6}
&AIPW&0.425 (0.172)&0.431 (-0.26, 0.94)&-0.005&0.172\\\cline{2-6}
&BR&0.432 (0.176)&0.437 (-0.27, 0.96)&0.002&0.176\\\hline
{\bf CBPS}&IPW&0.375 (0.132)&0.376 (-0.38, 0.74)&-0.055&0.143\\\cline{2-6}
&AIPW&0.413 (0.136)&0.418 (-0.36, 0.79)&-0.017&0.137\\\cline{2-6}
&BR&0.437 (0.145)&0.439 (-0.67, 0.84)&0.007&0.145\\\hline
{\bf bCART}&IPW&0.443 (0.115)&0.442 (0.09, 1.16)&0.013&0.116\\\cline{2-6}
&AIPW&0.423 (0.110)&0.424 (0.06, 1.09)&-0.007&0.110\\\cline{2-6}
&BR&0.491 (0.145)&0.485 (0.12, 1.65)&0.061&0.157\\\hline
{\bf Proposed}&IPW&0.458 (0.175)&0.455 (-0.10, 2.11)&0.028&0.178\\\cline{2-6}
&AIPW&0.387 (0.172)&0.388 (-0.21, 1.88)&-0.043&0.177\\\cline{2-6}
&BR&0.389 (0.176)&0.390 (-0.22, 1.99)&-0.041&0.181\\\hline
\end{tabular}
\end{center}
\end{table}

\begin{landscape}

\begin{figure}[h]
\begin{center}
\begin{tabular}{c}
\includegraphics[width=20cm]{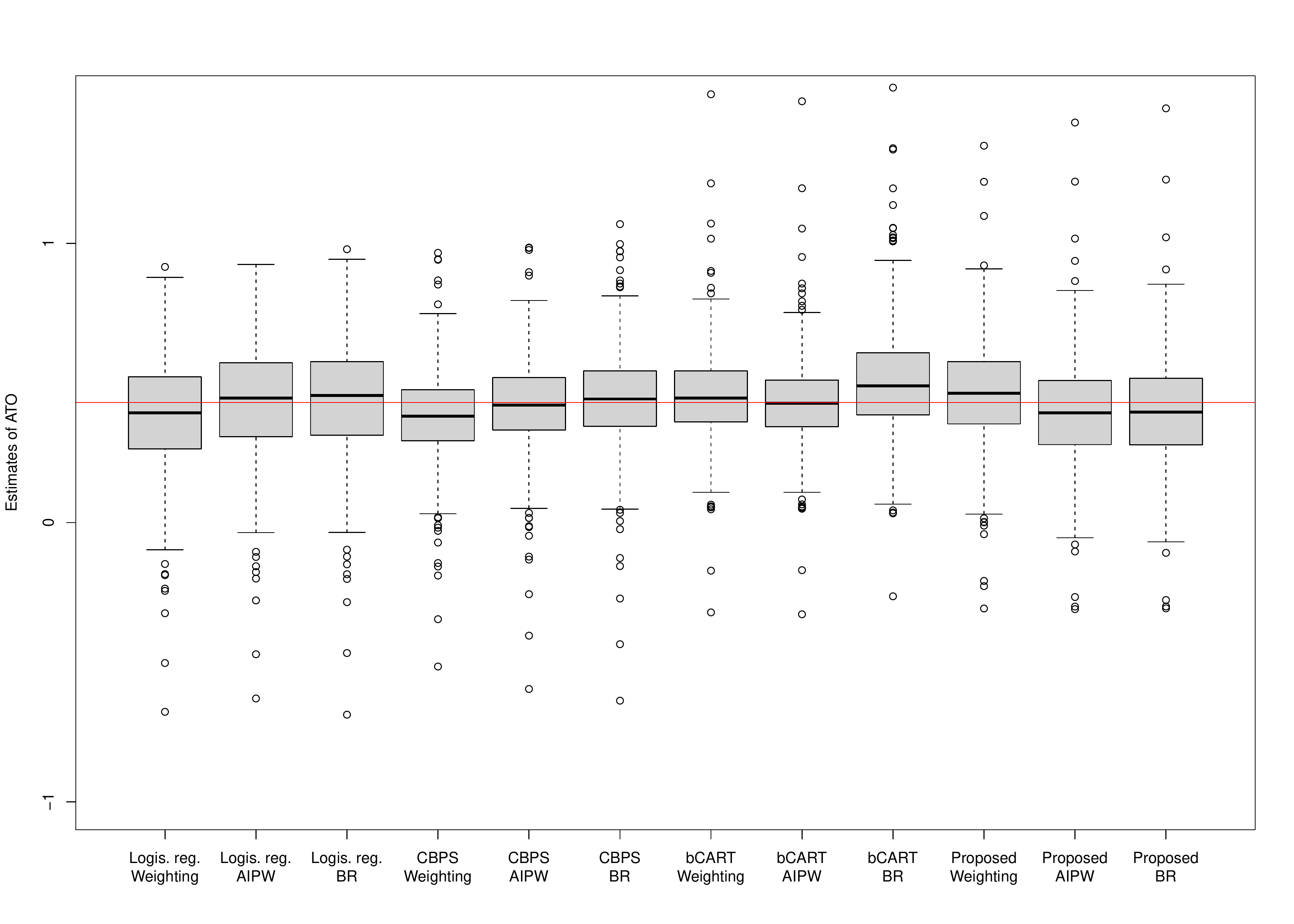}
\end{tabular}
\caption{Simulation results ($n=800$)}
\label{fig4}
\end{center}
\end{figure}

\begin{figure}[h]
\begin{center}
\begin{tabular}{c}
\includegraphics[width=20cm]{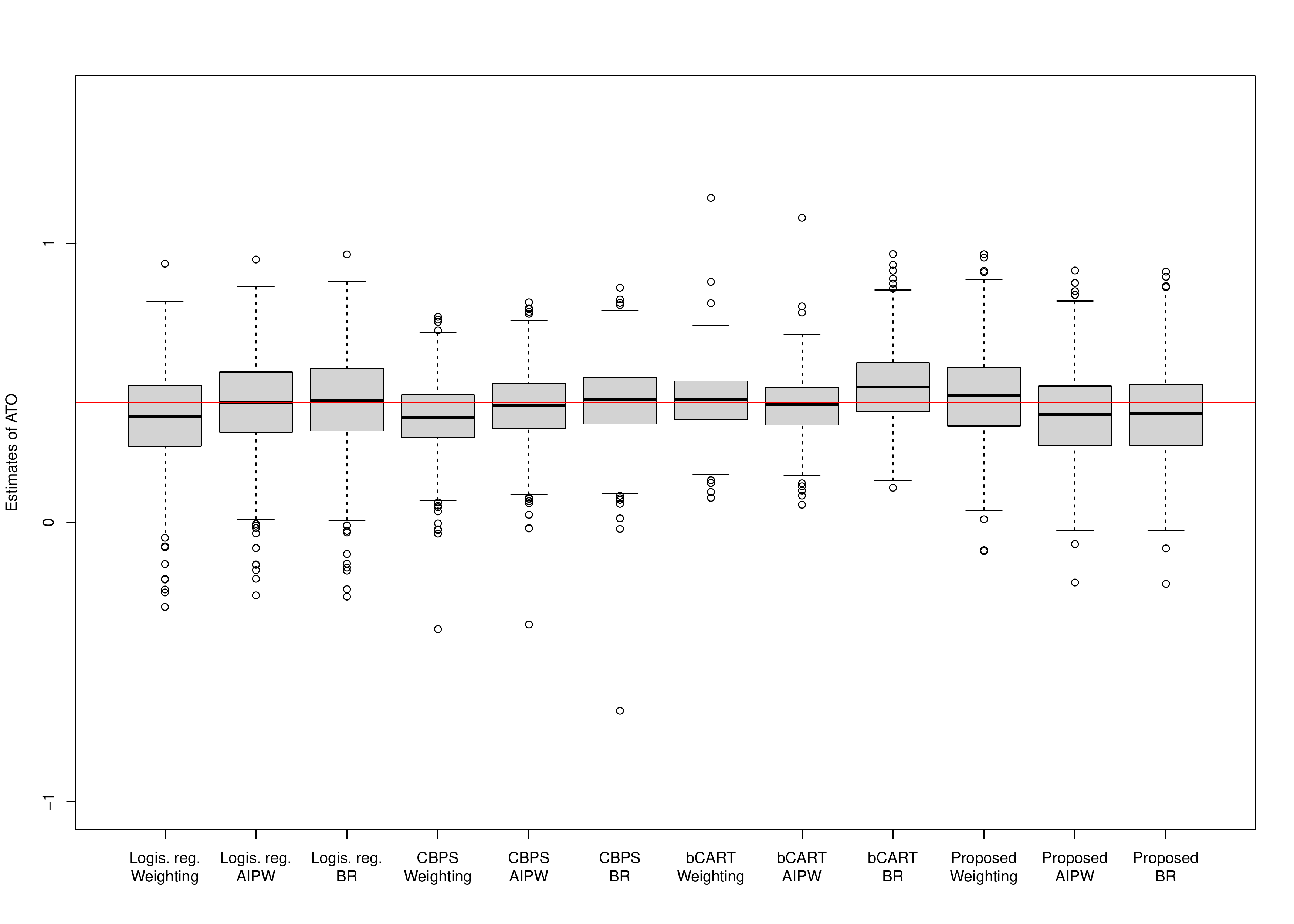}
\end{tabular}
\caption{Simulation results ($n=1200$)}
\label{fig5}
\end{center}
\end{figure}

\begin{figure}[h]
\begin{center}
\begin{tabular}{c}
\includegraphics[width=20cm]{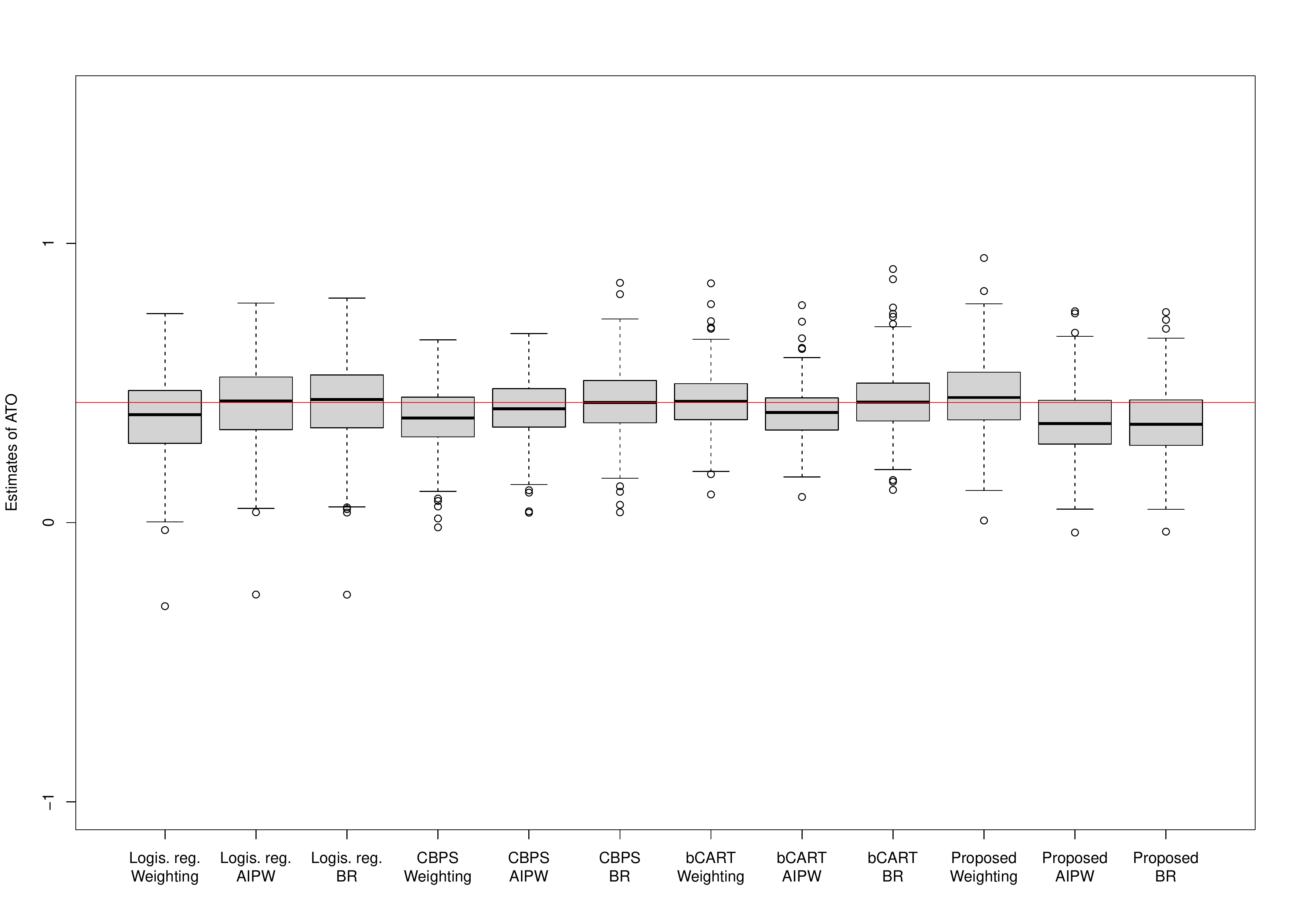}
\end{tabular}
\caption{Simulation results ($n=1600$)}
\label{fig6}
\end{center}
\end{figure}

\end{landscape}
\end{document}